\documentclass[aps,prd,preprint,nofootinbib,superscriptaddress]{revtex4}
\usepackage{color}
\usepackage{graphicx}
\newcommand{\uz}{[ {U}_0 ]}
\newcommand{\uzd}{[ {U}_0^\dagger ]}
\makeindex
\begin{document}
\preprint{OU-HET-823}
\title{Extracting the electro-magnetic pion form factor from QCD in a finite volume}
\author{Hidenori Fukaya}
\email[Email:]{hfukaya@het.phys.sci.osaka-u.ac.jp}

\author{Takashi Suzuki}
\email[Email:]{suzuki@het.phys.sci.osaka-u.ac.jp}

\affiliation{Department of Physics, Osaka University, Toyonaka, Osaka 560-0043 Japan}
%
%
%
\begin{abstract}
We consider finite volume effects on the electromagnetic form factor of the pion.  
We compute the peudoscalar-vector-pseudoscalar correlator 
in the $\epsilon$ expansion of chiral perturbation theory up to the next-to-leading order and 
find a way to remove the dominant part, which comes from a contribution of the pion zero-mode. 
Inserting non-zero momentum to relevant operators (or taking a subtraction of the correlators at different time-slices), 
and taking an appropriate ratio of them, one can automatically cancel 
the zero-mode's contribution, which becomes non-perturbatively large $\sim \mathcal{O}(100\%)$ in the $\epsilon$ regime.
The remaining finite volume dependence, which comes from the non-zero momentum modes,
is shown to be perturbatively small even in such an extremal case.
Since the zero-mode's dominance is universal in any finite volume scaling, 
and we do not rely on any particular feature of the $\epsilon$ expansion,
our method has a wide application to many other correlators of QCD.
\end{abstract}
%
%
\maketitle
%
%
%
%
%
%
\newpage
\section{Introduction}
\label{sec:introduction}

The electro-magnetic form factor of the charged pions
is one of the fundamental low-energy quantities in Quantum Chromodynamics (QCD). 
Experimentally, it is related to the pion charge radius $\langle r^2 \rangle_V$ 
through the relation
\begin{eqnarray}
\langle r^2 \rangle_V
&=&
\left. 6 \frac{d F_V(q^2)}{d q^2} \right|_{q^2 = 0},
\end{eqnarray}
where $F_V(q^2)$ denotes the electro-magnetic form factor at 
the momentum transfer $q^2$.
In terms of chiral perturbation theory (ChPT), 
it is related to the low-energy constant
$L_9$ (or $l_6$ in the $SU(2)$ case), 
which appears at the next-to-leading order (NLO) 
in the chiral Lagrangian \cite{Gasser:1983yg, Gasser:1984gg}.

However, it is still a non-trivial task for lattice QCD to fully reproduce or understand 
the low-energy behavior of the pion form factors. 
In fact, the lattice data of the pion charge radius have been sizably lower than the experimental value 
$\langle r^2 \rangle_V = 0.452 (11)~{\rm fm}^2$
(see the recent review in \cite{Brandt:2013ffb}). 
It is only recently that consistent values of $\langle r^2 \rangle_V$ were reported
by simulations near the physical point \cite{Koponen:Lat2013, JLQCD:Latt.2012PoS, Fukaya:2014jka}. 
According to ChPT, it is known that the pion charge radius shows 
a logarithmic divergence as the pion mass goes to zero. 
Thus, we may recognize that our simulated pion masses are too large to reproduce 
the logarithmic divergence, unless we directly simulate QCD near the chiral limit.
Namely, in order to examine the chiral logarithm of the pion charge radius,
it is essential to simulate lattice QCD in the very vicinity of the chiral limit.

Although current computational resources allow us to simulate QCD near the physical point,
one should carefully take two sources of systematic effects into account in such simulations.
One is the cut-off effects, especially those come from breaking of the chiral symmetry.
When the simulated quark mass is as small as the 
typical breaking scale of the chiral (flavor) symmetry 
(it is typically $\sim\Lambda_{\rm QCD}^3 a^2$ for the improved Wilson or staggered fermions, where
$\Lambda_{\rm QCD}$ is the QCD scale and $a$ denotes the lattice spacing),
it is known that the chiral logarithm is largely distorted. 
The low-lying Dirac eigenvalue spectrum, for example, is a quantity 
sensitive to such discretization effects \cite{Kieburg:2013xta}. 

Another source which may change the chiral behavior is the finite size of the lattice volume.
In the literature, it is often mentioned that the lattice size $L$ should satisfy $m_\pi L > 4$, 
where $m_\pi$ is a simulated pion mass \cite{Aoki:2013ldr},
to suppress the finite size effect at a few \% level.
Since the computational cost for inverting the Dirac operator increases as $m_\pi$ decreases,
it is demanding to keep $m_\pi L$ to be large enough.
Especially when we want to keep a good chiral symmetry 
to avoid the former discretization effects
on the chiral logarithm, and use a fermion formulation 
such as overlap or domain-wall fermions,
the available range of $m_\pi L$ is quite limited.

This naive criterion about $m_\pi L$, however, comes from the fact
that the zero-momentum mode of pions can propagate wrapping around
the lattice volume, whose contribution is typically 
given by $\exp(-m_\pi L)$. For the excited pion states,
the finite volume effects are much smaller, since
their discrete energy satisfies $E_\pi > 2\pi/L$ in a finite volume, and $E_\pi L>2\pi $.
Therefore, if we can eliminate or reduce 
the dominant contribution from the pion's zero-momentum mode,
one should be able to extract the low-energy quantities even on a small lattice.

In this work, we consider the ``worst'' case, the so-called $\epsilon$ regime of QCD,
to show that the above strategy actually works even in such an extremal situation.
In the $\epsilon$ regime, $m_\pi L < 1$, and the finite volume effects are
generally $\sim 100$\% and we receive a non-perturbatively large correction from the pion zero-mode.
However, using the $\epsilon$ expansion of ChPT \cite{Gasser:1986vb},
we compute the pseudoscalar-vector-pseudoscalar three-point function,
and find a way to automatically cancel the dominant part of them.
Since the zero-mode contribution has no space-time dependence,
two simple steps are enough to achieve this:
\begin{enumerate}
\item inserting non-zero momenta to relevant operators 
(or taking a subtraction of the correlators at different source points when one or two of the inserted momenta are zero).
\item taking ratios of them.
\end{enumerate}
We also compute the NLO corrections and show that these effects are actually 
suppressed by $1/F^2L^2$, where $F$ denotes the pion decay constant.
The preliminary result of this work has already appeared in Ref.~\cite{Fukaya-Suzuki}, 
and applied to numerical works by JLQCD collaboration \cite{JLQCD:Latt.2012PoS, Fukaya:2014jka}.

Here, we would like to remark the difference of our new approach
from the conventional ones in the $\epsilon$ regime.
In the previous works, the $\epsilon$ expansion was used
to disentangle the low-energy constants \cite{Hansen:1990un, Hansen:1990yg, Hernandez:2002ds, Giusti:2004an, Bernardoni:2008ei, Hernandez:2008ft, Aoki:2011pza}, 
using a bunch of Bessel functions,
from the lattice data which were largely contaminated by the finite volume effects.
In this work, we use (the $\epsilon$ expansion of) ChPT
in more indirect way : just for finding 
the combination of the correlators 
which has a small sensitivity to the volume.
As we will see in the following sections, this idea makes the analysis 
in the $\epsilon$ regime of QCD greatly simplified.
In particular, we would like to emphasize that 
there is essentially no need to use Bessel functions
for the computation of  the pion form factor.
Moreover, since the dominance of the pion zero-mode's contribution
(having the longest correlation length), is universal
for any finite volume effects on any operators,
we expect a wide application of this method.
It may be useful for heavier hadron form factors, 
and simulations in the $p$ regime as well.

The rest of this paper is organized as follows. 
In Sec.~\ref{sec:chpt}, we review the $\epsilon$ expansion of ChPT and present how to compute the correlators at one-loop level.
In Sec.~\ref{sec:twopoint}, we consider the two-point functions to illustrate our new idea.
Then, our main result for the pseudoscalar-vector-pseudoscalar three-point functions is presented in Sec.~\ref{sec:threepoint},  
including the NLO effects.
In Sec.~\ref{sec:extraction}, we show how to extract the pion vector form factor, 
and estimate the remaining finite volume effects numerically : 
we find that it is a few percent level already at $L=3~{\rm fm}$. 
Summary and conclusion are given in Sec.~\ref{sec:summary}.


\section{The $\epsilon$ expansion of ChPT}
\label{sec:chpt}
In this section, we review the $\epsilon$ expansion of ChPT,
and show how to perform the one-loop level calculation of the correlators.
First, we give the counting rule of the $\epsilon$ expansion. Second, we write down 
the chiral lagrangian with pseudo-scalar and vector source terms, 
and explain a general procedure to calculate correlators from a partition function. 
Finally, we give the technical details of this study at the end of this section.

\subsection{The chiral Lagrangian}
We consider $N_f$-flavor ChPT in an Euclidean finite volume $V= TL^3$ with the periodic boundary condition in every direction.
The Lagrangian \cite{Gasser:1983yg, Gasser:1984gg} is given by
\begin{eqnarray}
\label{eq:ChL}
\mathcal{L}_{\rm ChPT}
=
\frac{F^2}{4} {\rm Tr} \left[ \left( \partial_\mu U(x) \right)^\dagger \left( \partial_\mu U(x) \right) \right]
-
\frac{\Sigma}{2} {\rm Tr} \left[ \mathcal{M}^\dagger U(x) + U^\dagger (x) \mathcal{M} \right]
+
\cdots,
\end{eqnarray}
where $U(x)$ denote the chiral field which is an element of the group $SU(N_f)$. 
$\Sigma$ is the chiral condensate and $F$ is the pion decay constant both in the chiral limit. 
The terms omitted by ellipses are the ones at the higher orders. 
For simplicity, we take the quark mass matrix $\mathcal{M}$ degenerate and diagonal:
$\mathcal{M} = {\rm diag} (m,m,m,\cdots)$.

In the $\epsilon$ regime \cite{Gasser:1986vb}, the vacuum is not fixed 
but has non-perturbatively large fluctuations. 
Namely, the zero-mode of the pions must be integrated exactly. 
Thus, we separate it from the non-zero momentum modes and parametrize the chiral field as 
\begin{eqnarray}
\label{eq:eexp}
U(x)
=
U_0 \, \left( \frac{i \sqrt2 }{F} \xi(x) \right),
\hspace{10pt}
U_0 \in SU(N_f),
\end{eqnarray}
where $U_0$ denotes the zero-modes. The non-zero momentum mode is decomposed as 
$\xi (x) = T^a \xi^a(x)$ with $SU(N_f)$ generators $T^a$, 
for which we use the normalization of ${\rm Tr} [T^a T^b] = \frac{1}{2}\delta^{ab}$. 
Since the constant modes are separated from $\xi (x)$ fields as $U_0$, a constraint
\begin{eqnarray}
\label{eq:constraint}
\int d^4x\, \xi (x) = 0,
\end{eqnarray}
must be satisfied to avoid  the double-counting of the zero-modes.

Now, we rewrite the chiral Lagrangian Eq.~(\ref{eq:ChL}) 
with the $\epsilon$ expansion, whose counting rule is given by
\begin{eqnarray}
\label{eq:ecount}
&&U_0 \sim \mathcal{O} (1),
\nonumber \\
&&\epsilon \sim \partial_\mu \sim \frac{1}{V^{1/4}} \sim m_\pi^{1/2} \sim m^{1/4} \sim \xi(x),
\end{eqnarray}
as
\begin{eqnarray}
\mathcal{L}_{\rm ChPT}
&=&
-\frac{\Sigma}{2} {\rm Tr} \left[ \mathcal{M}^\dagger U_0 + U_0^\dagger \mathcal{M} \right]
+
\frac{1}{2} {\rm Tr} \left[ \partial_\mu \xi \partial_\mu \xi \right] (x)
\nonumber \\ &&
+
\frac{\Sigma}{2F^2} {\rm Tr} \left[ \left( \mathcal{M}^\dagger U_0 + U_0^\dagger \mathcal{M} \right) \xi^2 \right] (x)
+
\cdots.
\end{eqnarray}
From this Lagrangian, one can recognize that we are treating a hybrid system 
containing bosonic $\xi (x)$ fields and a matrix $U_0$, which are weakly interacting.

For $\xi(x)$ fields, one can perform the Gaussian integrals without difficulty.
In this work, we use the correlator in quark-line basis,
\begin{eqnarray}
\label{eq:xipropagator}
\langle [\xi(x)]_{ij} [\xi (y)]_{kl} \rangle_\xi
&=&
\delta_{il} \delta_{jk} \bar{\Delta}(x-y) - \delta_{ij} \delta_{kl} \frac{1}{N_f} \bar{\Delta}(x-y),
\end{eqnarray}
where the second term comes from the constraint ${\rm Tr}\xi =0$, and
\begin{eqnarray}
\bar{\Delta} (x)
&\equiv&
\frac{1}{V} \sum_{p \neq 0} \frac{e^{ipx}}{p^2},
\end{eqnarray}
describes the propagation of the massless bosons. Here, the summation is taken over the non-zero 4-momentum
$p = 2 \pi (n_t/T, n_x/L, n_y/L, n_z/L),$
with integers $n_\mu$, except for $p=(0,0,0,0)$, 
because of the constraint Eq.~(\ref{eq:constraint}).

While $\xi (x)$ fields are treated perturbatively, the zero-mode denoted by $U_0$ 
has to be non-perturbatively integrated (we will denote it by $ \langle \cdots \rangle_{U_0}$). 
It is known that these matrix integrals are expressed by
the Bessel functions \cite{Leutwyler:1992yt, Splittorff-Verbaarschot, Fyodorov-Akemann}, which is
a peculiar feature of the $\epsilon$ regime.
Historically, this special feature of the $\epsilon$ regime
is used for extracting the leading LEC's, $\Sigma$ and $F$,
which are more sensitive to the volume than others.
However, for the other LEC's at NLO, we should take a different strategy, or
we should remove the contamination from the finite size.
In this work on the vector form factor of pions, which is related to $L_9$,
the $U_0$ integral plays a less important role.

\subsection{Partition function and correltors}
In this subsection, we consider the partition function 
of ChPT in the $\epsilon$ regime and show how
to calculate the correlation functions.
First, we introduce the relevant source terms to the chiral Lagrangian Eq.~(\ref{eq:ChL}). 
Since the Lagrangian is invariant under the chiral rotation,
\begin{eqnarray}
U(x) \to g_L U(x) g_R^\dagger,
\hspace{10pt}
g_L, g_R \in SU(N_f),
\end{eqnarray}
the vector or axial vector operators are given through the Noether's theorem 
for the vectorlike transformation $g_L =g_R$ and the axial one $g_L =g_R^\dagger$.
It is easy to see that adding these operators
is equivalent to replacing the derivatives by the ``covariant'' derivatives:
\begin{eqnarray}
\partial_\mu
\to
\nabla_\mu U(x) \equiv \partial_\mu U(x)
-
i(v_\mu (x) + a_\mu (x)) U(x)
+
i U(x) (v_\mu (x) - a_\mu (x)),
\end{eqnarray}
where $v_\mu (x)$ and $a_\mu (x)$ denote the vector and axial-vector sources, respectively.
Similarly, since the Lagrangian is invariant under the Parity transformation,
\begin{eqnarray}
U(x) \to U^\dagger (x),
\hspace{10pt}
x = (t, x, y, z) \to x = (t, -x, -y, -z),
\end{eqnarray}
adding a scalar $U(x) + U^\dagger (x)$ 
and a pseudoscalar $U(x) - U^\dagger (x)$ is absorbed in the mass matrix:
\begin{eqnarray}
\mathcal{M} \to \mathcal{M}_J \equiv \mathcal{M} + s(x) + ip(x),
\end{eqnarray}
where $s(x)$ and $p(x)$ denote the scalar and pseudoscalar sources, respectively.
we set $s(x)=a_\mu(x)=0$ in the following.

Next, let us introduce the NLO terms of the chiral lagrangian.
However, some of them are irrelevant to our calculations.
In this study, it is enough to consider the terms with
the low-energy constants $L_i$ ($i=4,\cdots 9$).
Namely, we consider the Lagrangian 
\begin{eqnarray}
\mathcal{L} (s,p,v_\mu,a_\mu)
&=&
\frac{F^2}{4}{\rm Tr}
[\nabla_\mu U^\dagger(x)\nabla_\mu U(x) ]
-\frac{\Sigma}{2}{\rm Tr}
[\mathcal{M}_J^{\dagger}
U(x)
+U^\dagger(x) 
\mathcal{M}_U]\nonumber\\
&&+L_4\frac{2\Sigma}{F^2}{\rm Tr}[(\nabla_\mu U(x))^\dagger\nabla_\mu U(x) ]
\times {\rm Tr}
[\mathcal{M}_J^{\dagger} U(x)
+U^\dagger(x)\mathcal{M}_J]
\nonumber\\
&&+L_5\frac{2\Sigma}{F^2}
{\rm Tr}\left[(\nabla_\mu U(x))^\dagger\nabla_\mu U(x) 
(\mathcal{M}_J^{\dagger} U(x)
+U^\dagger(x)\mathcal{M}_J)\right]
\nonumber\\
&&-L_6 \left(\frac{2\Sigma}{F^2}
{\rm Tr}
[\mathcal{M}_J^{\dagger} U(x)
+U^\dagger(x)\mathcal{M}_J]\right)^2
\nonumber\\
&&-L_7 \left(\frac{2\Sigma}{F^2}
{\rm Tr}
[\mathcal{M}_J^{\dagger} U(x)
-U^\dagger(x)\mathcal{M}_J]\right)^2
\nonumber\\
&&-L_8 \left(\frac{2\Sigma}{F^2}\right)^2
{\rm Tr}
[\mathcal{M}_J^{\dagger} U(x)
\mathcal{M}_J^{\dagger} U(x)
+U^\dagger(x)\mathcal{M}_J
U^\dagger(x)\mathcal{M}_J]
\nonumber\\
&&+iL_9 
{\rm Tr}
[F_{\mu\nu}^R (x)\nabla^\mu U(x)
(\nabla^\nu U(x))^\dagger
+F_{\mu\nu}^L (x)(\nabla^\mu U(x))^\dagger \nabla^\nu U(x)],
\end{eqnarray}
where 
\begin{eqnarray}
F_{\mu\nu}^I(x) &=& \partial_\mu F^I_\nu(x)-\partial_\nu F^I_\mu(x)
-i[F^I_\mu(x), F^I_\nu(x)], \;\;\;\;\; I=R,L,\nonumber\\
F^R_\mu(x)&=&v_\mu(x)+a_\mu(x), \;\;\;\;\;F^L_\mu(x)=v_\mu(x)-a_\mu(x).
\end{eqnarray}

The calculation of ChPT is performed in the functional
integral formalism.
The partition function is defined by
\begin{eqnarray}
\mathcal{Z}(s,p,v_\mu,a_\mu) = \int \prod_x d U(x) 
\exp\left[-\int d^4x \mathcal{L}(s,p,v_\mu,a_\mu) \right],
\end{eqnarray}
and the correlators are computed by differentiating
it with respect to the corresponding sources, and take their zero limits.
The pseudoscalar two-point function, for example, is given by
\begin{eqnarray}
\langle P^a(x)P^b(y) \rangle 
&=& \left.\frac{1}{\mathcal{Z}(0,0,0,0)}\frac{\delta}{\delta p^a(x)}
\frac{\delta}{\delta p^b(y)}\mathcal{Z}(s,p,v_\mu,a_\mu)
\right|_{s,p,v_\mu,a_\mu =0},
\end{eqnarray}
where $p^a(x)$ denotes the coefficient of 
an $SU(N_f)$ generator $T^a$, where we decompose 
the source as $p(x)=T^a p^a(x)$.

One should note that our non-trivial parametrization of $U(x)$
needs a non-trivial Jacobian in the functional integration measure:
\begin{eqnarray}
\int \prod_x dU(x) &=& \int dU_0 \prod_x d\xi(x) \mathcal{J}(U_0,\xi).
\end{eqnarray}
A perturbative calculation \cite{Hansen:1990un, Bernardoni:2007hi}
has shown
\begin{eqnarray}
\label{eq:measure}
\mathcal{J}(U_0,\xi)&=&\exp\left(-\int d^4x \frac{N_f}{3F^2V}{\rm Tr}\;\xi^2(x)
+{\cal O}(\epsilon^4)\right),
\end{eqnarray}
which can be regarded as an additional mass term of the 
$\xi(x)$ fields at the one-loop level.
Note that this additional mass doesn't vanish even in the 
$m\to 0$ limit, which keeps the theory infra-red finite.

Finally, let us consider the $\theta$ vacuum and fixing topology. 
In the $\epsilon$ regime, we often consider a fixed topological sector, 
rather than the full QCD vacuum with the vacuum angle $\theta=0$. 
For this purpose, we encode the non-zero vacuum angle $\theta$ to the mass term \cite{Aoki:2009mx}, 
\begin{eqnarray}
\mathcal{M}\to \mathcal{M}_\theta = \mathcal{M}\exp(-i\theta/N_f),
\end{eqnarray}
using the axial $U(1)_A$ rotation. 
Then we can perform a Fourier transformation with respect to $\theta$
to obtain the partition function at fixed topology,
\begin{eqnarray}
\mathcal{Z}^Q (s,p,v_\mu,a_\mu) \equiv \int_0^{2\pi} \frac{d\theta}{2\pi} 
\left[\left. e^{i\theta Q} 
\mathcal{Z}(s,p,v_\mu,a_\mu)\right|_{\mathcal{M}=\mathcal{M}_\theta}\right],
\end{eqnarray}
where $Q$ denotes the topological charge of the original gauge fields. It is known that this $\theta$ integral can be absorbed in the group integration of the zero-mode: redefining the zero mode,
\begin{eqnarray}
\label{eq:eexpfixedQ}
e^{i\theta/N_f}U(x) &=& \overline{U}_0\exp\left(\frac{i\sqrt{2}}{F}\xi(x)\right),
\end{eqnarray}
where $\overline{U_0} \in U(N_f)$,
the zero-mode part of the functional integral is modified to
\begin{eqnarray}
\int \frac{d\theta}{2\pi}\exp(i\theta Q)\int_{SU(N_f)} dU_0 
F(\mathcal{M}^\dagger e^{i\theta/N_f} U_0) 
= \int_{U(N_f)} d\overline{U}_0 (\det\overline{U}_0)^Q 
F(\mathcal{M}^\dagger\overline{U}_0),  
\end{eqnarray}
where we have used the fact that 
the zero-mode in the Lagrangian always appears
as a function of $\mathcal{M}^\dagger e^{i\theta/N_f} U_0$ 
(and its Hermitian conjugate).
Fixing the topology is technically easier since
the $U(N_f)$ group integral is simpler than that of $SU(N_f)$.
It is also useful for investigating the finite volume physics
which is sensitive to the topology of the gauge fields.
It is important to note that the fixing topology effect is
totally encoded in the pion-zero mode, and therefore,
is automatically eliminated once the effect of the latter is eliminated.
Since we will be able to cancel the effect of $U_0$ (from the LO contribution),
in the following sections, we don't distinguish 
$U_0$ and $\overline{U}_0$ unless explicitly stated.

We are now ready for the 1-loop computations.
However, we would like to give some useful 
technical details which simplify the calculations, 
in the next subsection.

\subsection{Technical details}\label{sec:technicaldetails}
Because of the non-trivial parametrization of the chiral field,
we have a lot of diagrams to be computed in the $\epsilon$ expansion of ChPT even at NLO.
Here we rewrite the Lagrangian using the non-self-contracting (NSC) vertices,
and compute some of 1-loop diagrams in advance, as corrections to the chiral Lagrangian. 
This reduces the number of diagrams and simplify our calculation.

The $n$-point NSC vertex is defined by
\begin{eqnarray}
\label{eq:nptNSC}
[\xi^n(x)]^{NSC} &\equiv&  \xi^n(x)
- (\mbox{all possible $\xi$ contractions}).
\end{eqnarray}
and we can absorb the contracted part in the redefinition 
of the lower dimensional terms in the Lagrangian.
Note that $\langle [\xi^n(x)]^{NSC}\rangle_\xi=0$ by definition.
For example, a term in the Lagrangian at NLO can be re-expressed by
\begin{eqnarray}
\frac{1}{6F^2}{\rm Tr}
[\partial_\mu \xi \xi \partial_\mu \xi \xi-\xi^2(\partial_\mu \xi)^2]
&=& \frac{1}{6F^2}{\rm Tr}
[\partial_\mu \xi \xi \partial_\mu \xi \xi-\xi^2(\partial_\mu \xi)^2]^{NSC}
\nonumber\\&&
+\frac{1}{2}{\rm Tr}\left[
(\partial_\mu \xi)^2\right]\Delta Z^{\xi}
+\frac{1}{2}{\rm Tr}\left[\xi^2
\right]\Delta M^2,
\end{eqnarray}
where
\begin{eqnarray}
\Delta M^2 &=& - \frac{N_f}{3F^2} \partial_\mu^2\bar{\Delta}(0)
 = \frac{N_f}{3F^2 V},
\end{eqnarray}
can be absorbed in the re-definition of the mass term, and
\begin{eqnarray}
\Delta Z^\xi &=& - \frac{N_f}{3F^2} \bar{\Delta}(0),
\end{eqnarray}
can be absorbed in the re-definition of the kinetic term.
Here, and in the following, the momentum summations in 
embeded in $\bar{\Delta}(0)$ etc. 
are kept unperformed until the very end of the calculation,
except for the trivially clear cases like $\partial_\mu^2\bar{\Delta}(0)=-1/V$,
$\partial_\mu\bar{\Delta}(0)=0$. 
In this work, we employ the dimensional regularization for the loop integrals.

With the NSC vertices, the action is expanded as
\begin{eqnarray}
\mathcal{S}_{\rm ChPT} &=& \int d^4 x\, \mathcal{L}
= \mathcal{S}^{\rm LO} +\mathcal{S}^{\rm NLO} + \mathcal{S}^{\rm src}+\cdots,
\end{eqnarray}
where
\begin{eqnarray}
\mathcal{S}^{\rm LO} &=& 
-\frac{Z^{\Sigma}\Sigma V}{2}{\rm Tr}
[\mathcal{M}^{\dagger}
U_0+U_0^\dagger 
\mathcal{M}]
+\int d^4 x \left\{ 
\frac{1}{2}{\rm Tr}[\partial_\mu \xi\partial^\mu \xi](x)
\right\}(Z^\xi)^2,
\\
\label{eq:SNLOeff}
\mathcal{S}^{\rm NLO} &=& \mathcal{S}^{\rm NLO}_{K} + \mathcal{S}^{\rm NLO}_M,
\nonumber\\
\mathcal{S}^{\rm src} &=& \int d^4 x 
{\rm Tr}\left[p(x)P(x)+v_\mu (x)V^\mu (x)\right],
\end{eqnarray}
where
\begin{eqnarray}
\mathcal{S}^{\rm NLO}_{K} &\equiv&
\int d^4 x 
\frac{1}{6 F^2}{\rm Tr}[\partial_\mu \xi\xi\partial^\mu \xi\xi
-\xi^2 \partial_\mu \xi\partial^\mu \xi]^{NSC}(x),\\
\mathcal{S}^{\rm NLO}_M &\equiv&
\int d^4 x\frac{\Sigma}{2F^2}{\rm Tr}
\left[\left(\mathcal{M}^{\dagger}
U_0+U_0^\dagger \mathcal{M}+\frac{N_f}{\Sigma V}\right)\xi^2\right]^{NSC}(x).
\end{eqnarray}
Note that the linear term in $\xi(x)$ disappears because of the constraint Eq.~(\ref{eq:constraint}).

Here, the source operators are given by 
\begin{eqnarray}
\label{eq:Pnsc}
P(x) &=& iZ^{P1}\frac{\Sigma}{2}[U_0-U_0^\dagger]
-Z^{P2}\frac{\Sigma}{\sqrt{2}F}[U_0\xi+\xi U_0^\dagger]
-Z^{P3}\frac{i\Sigma}{2F^2}[U_0\xi^2-\xi^2U_0^\dagger]^{NSC}
\nonumber\\&&
+\frac{i\Sigma}{12F^4}\bar{\Delta}(0)[U_0-U_0^\dagger]{\rm Tr}[\xi^2]^{NSC}
\nonumber\\&&
+\frac{\Sigma}{3\sqrt{2}F^3}[U_0\xi^3+\xi^3U_0^\dagger]^{NSC}
+\frac{i\Sigma}{12F^4}[U_0\xi^4-\xi^4U_0^\dagger]^{NSC}
\nonumber\\&&
-iL_4\frac{4\Sigma}{F^4}\left({\rm Tr}[\partial_\mu \xi\partial^\mu \xi]^{NSC}\right)
\times[U_0-U_0^\dagger]
-iL_5\frac{4\Sigma}{F^4}[U_0 \partial_\mu \xi\partial^\mu \xi 
-\partial_\mu \xi\partial^\mu \xi  U_0^\dagger]^{NSC}
\nonumber\\&&
+{\cal O}(\epsilon^5),\\
\label{eq:V}
V^\mu (x) &=& 
-\frac{FZ^{V1}}{\sqrt{2}}\left[
U_0\partial^\mu \xi U_0^\dagger -\partial^\mu \xi \right]
\nonumber\\&&
+\frac{iZ^{V2}}{2}\left[
U_0(\partial^\mu \xi\xi-\xi\partial^\mu \xi) U_0^\dagger 
+(\partial^\mu \xi\xi-\xi\partial^\mu \xi)
\right]^{NSC}
\nonumber\\&&
+\frac{1}{3\sqrt{2}F}\left[
U_0 (\partial^\mu \xi\xi^2-2\xi\partial^\mu \xi \xi+\xi^2 \partial^\mu \xi)U_0^\dagger 
- (\partial^\mu \xi\xi^2-2\xi\partial^\mu \xi \xi+\xi^2 \partial^\mu \xi)
\right]^{NSC}
\nonumber\\&&
-\frac{i}{12F^2}\left[
U_0(\partial^\mu \xi\xi^3-3\xi\partial^\mu \xi\xi^2+3\xi^2\partial^\mu\xi \xi-\xi^3 \partial^\mu \xi) U_0^\dagger 
\right.\nonumber\\&&\left.\hspace{1in}
+(\partial^\mu \xi\xi^3-3\xi\partial^\mu \xi\xi^2+3\xi^2\partial^\mu \xi\xi-\xi^3 \partial^\mu \xi)
\right]^{NSC}
\nonumber\\&&
-\frac{2iL_9}{F^2}\partial_\nu \left[
U_0(\partial^\nu \xi\partial^\mu\xi-\partial^\mu \xi\partial^\nu\xi) U_0^\dagger 
+(\partial^\nu \xi\partial^\mu\xi-\partial^\mu \xi\partial^\nu\xi)
\right]^{NSC}+{\cal O}(\epsilon^6),
\end{eqnarray}
where
\begin{eqnarray}
Z^\Sigma &=& 1-\frac{N_f^2-1}{N_fF^2}\bar{\Delta}(0),\\
Z^\xi &=& 1-\frac{N_f}{6F^2}\bar{\Delta}(0),\\
Z^{P1}&=& Z^\Sigma + {\cal O}(\epsilon^4),\\
Z^{P2}&=& 1-\frac{2N_f^2-3}{3N_fF^2}\bar{\Delta}(0),\\
Z^{P3}&=& 1-\frac{N_f}{2F^2}\bar{\Delta}(0)+\frac{1}{N_f F^2}\bar{\Delta}(0),\\
Z^{V1}&=& 1-\frac{2N_f}{3F^2}\bar{\Delta}(0),\\
Z^{V2}&=& 1-\frac{5N_f}{6F^2}\bar{\Delta}(0).
\end{eqnarray}
In the above expression, the argument $(x)$ of $\xi(x)$ is omitted for simplicity.
In this work, we don't consider contact correlators at the same position, such as $\langle P(x)V(x) \rangle$.
We have, therefore, only collected the  terms linear in the sources $p(x)$ and $v_\mu (x)$.

Here we note that except for $Z^{V2}$, we can absorb
all the $Z$ factors into the redefinition of the
wave functions ($\xi$ fields), or the coupling constants,
by defining
\begin{eqnarray}
\xi^\prime (x)&\equiv& Z^\xi\xi(x),\\
\label{eq:zfunctions2}
\Sigma_{\rm eff} &\equiv& Z^\Sigma \Sigma,\\
\label{eq:zfunctions3}
F_{\rm eff} &\equiv& F\left(1-\frac{N_f}{2F^2}\bar{\Delta}(0)\right).
\end{eqnarray}
Therefore, except for the 4-th term in Eq.~(\ref{eq:Pnsc}),
the vertex corrections of the
two-point and three-point correlators can be
obtained by simply replacing the coefficients of 
the LO results with the shifted ones $\Sigma_{\rm eff}$ and $F_{\rm eff}$,
except for multiplying the coefficient of the
second term in $V^\mu (x)$,
\begin{eqnarray}
\label{eq:zv2}
Z^{V2}/(Z^{\xi})^2 &=& 1-\frac{N_f}{2F^2}\bar{\Delta}(0),
\end{eqnarray}
and the third term in $P(x)$,
\begin{eqnarray}
\label{eq:zp3}
Z^{P3^\prime}\equiv\frac{Z^{P3}}{Z^\Sigma(Z^{\xi})^2}\left(\frac{F_{\rm eff}}{F}\right)^2 
&=& 1-\frac{N_f}{6F^2}\bar{\Delta}(0).
\end{eqnarray}

With this action, for any operator $O$ (as a function of $\xi$ and $U_0$) 
in the $\epsilon$ expansion,
\begin{eqnarray}
O &=& O^{\rm LO} + O^{\rm NLO} + \cdots,
\end{eqnarray}
its expectation value is perturbatively evaluated as,
\begin{eqnarray}
\label{eq:vev}
\langle O \rangle &\equiv& 
\frac{\displaystyle \int \mathcal{D}U_0 \mathcal{D}\xi \left[\;
(O^{\rm LO} + O^{\rm NLO} + \cdots) e^{-\mathcal{S}^{\rm LO}-\mathcal{S}^{\rm NLO}+\cdots}\right]}
{\displaystyle \int \mathcal{D}U_0 \mathcal{D}\xi\; \left[e^{-\mathcal{S}^{\rm LO}-\mathcal{S}^{\rm NLO}+\cdots}\right]}
\nonumber\\&=&
\langle \langle O^{\rm LO} \rangle_\xi \rangle_{U_0}
+\left[\langle \langle O^{\rm NLO} \rangle_\xi \rangle_{U_0}
- \langle \langle O^{\rm LO} \mathcal{S}^{\rm NLO}\rangle_\xi \rangle_{U_0}
+ \langle \langle O^{\rm LO} \rangle_\xi \rangle_{U_0}\langle \langle \mathcal{S}^{\rm NLO}\rangle_\xi \rangle_{U_0}
\right]
+\cdots ,\nonumber\\
\end{eqnarray}
where we have used the following notations,
\begin{eqnarray}
\label{eq:zeromodeintegral}
\langle O_1(U_0)\rangle_{U_0} 
&\equiv& 
\frac{\displaystyle\int \mathcal{D} U_0\;e^{\frac{\Sigma_{\rm eff} V}{2}{\rm Tr}
[\mathcal{M}^\dagger U_0+U_0^\dagger \mathcal{M}]}\;O_1(U_0)
}
{\displaystyle\int \mathcal{D}U_0\;e^{\frac{\Sigma_{\rm eff} V}{2}{\rm Tr}
[\mathcal{M}^\dagger U_0+U_0^\dagger \mathcal{M}]} },\\
\langle O_2(\xi)\rangle_\xi &\equiv& 
\frac{\displaystyle\int \mathcal{D}\xi\; e^{-\int d^4x 
\frac{1}{2}{\rm Tr}[\xi(-\partial_\mu^2)\xi](x)}
O_2(\xi)}
{\displaystyle\int \mathcal{D}\xi\; e^{-\int d^4x 
\frac{1}{2}{\rm Tr}[\xi(-\partial_\mu^2)\xi](x)}}.
\end{eqnarray}
Note that, due to the use of NSC vertices, we don't need to calculate 
the fourth term in Eq.~(\ref{eq:vev})
since $\langle \mathcal{S}^{\rm NLO}\rangle_\xi =0$.

In the usual $\theta =0$ vacuum, $\mathcal{D} U_0$ 
denotes a Haar measure on $SU(N_f)$, while
it should be replaced by $\mathcal{D} U_0 (\det U_0)^Q$
on $U(N_f)$, for a fixed topological sector as discussed in the previous subsection.

\section{Two-point functions}
\label{sec:twopoint}

As we have mentioned in Sec.~\ref{sec:introduction}, 
the dominant finite volume effect on correlators comes from the pion zero-mode. 
Since the zero-mode itself does not depend on the space-time position $x$, 
its effect always appear as an $x$-independent constant term or overall constants
of $x$-dependent terms.
In either case, it is not difficult to eliminate these zero-mode's effects
from the correlators.
In this section, we demonstrate this new idea 
taking the two-point pseudoscalar correlators, as an easiest example.

\subsection{LO calculation}
Let us consider a pseudoscalar operator in the charged pion channel,
\begin{eqnarray}
P^1(x)
\equiv
\frac{1}{2}\left( \left[ P(x) \right]_{12} + \left[ P(x) \right]_{21} \right).
\end{eqnarray}
From the chiral symmetry, it is easy to confirm that
its two-point function satisfies
\begin{eqnarray}
\langle [P(x)]_{12}[P(y)]_{12} \rangle &=& \langle [P(x)]_{21}[P(y)]_{21} \rangle=0,
\end{eqnarray}
and 
\begin{eqnarray}
\label{eq:simplify}
\langle  \left[ P(x) \right]_{12} \left[ P(y) \right]_{21}\rangle
&=& \langle \left[ P(x) \right]_{21} \left[ P(y) \right]_{12}\rangle =
2\langle  P^1(x) P^1(y)\rangle.
\end{eqnarray}
The quark field basis $ \left[ P(x) \right]_{ij}$ is convenient
unless we consider the neutral sector of ChPT, 
since  $\langle  \left[ P(x) \right]_{ij} \left[ P(y) \right]_{ji}\rangle$
shares the same normalization of the so-called ``connected'' 
contribution of the conventional meson correlators in lattice QCD.
Therefore, we use $ \left[ P(x) \right]_{ij}$
rather than the original $P^1(x)$ in the following analysis.

Now we can write down the two-point function to ${\cal O}(\epsilon^2)$,
\begin{eqnarray}
\label{eq:ppcorrelator}
\langle [P(x)]_{12}[P(y)]_{21}\rangle
&=&
-\frac{\Sigma_{\rm eff}^2}{4}
\left\langle \mathcal{A}(U_0) \right\rangle_{U_0}
+
\frac{\Sigma_{\rm eff}^2}{2 F_{\rm eff}^2 V}
\left\langle \mathcal{B}(U_0)\right\rangle_{U_0}
\sum_{p\neq0} \frac{e^{ip(x-y)}}{p^2},
\end{eqnarray}
where
\begin{eqnarray}
\mathcal{A} (U_0)
&=&
[U_0-U_0^\dagger]_{12} [U_0-U_0^\dagger]_{21}
+
\frac{1}{2}([U_0-U_0^\dagger]_{12})^2
+
\frac{1}{2}([U_0-U_0^\dagger]_{21})^2,\\
\label{eq:simpleAB}
\mathcal{B}(U_0) &=& 
2 + [U_0]_{11}[U_0]_{22}+[U_0^\dagger]_{11}[U_0^\dagger]_{22}
-\left([U_0]_{12}+[U_0^\dagger]_{12})([U_0]_{21}+[U_0^\dagger]_{21}\right)/N_f.
\label{eq:simpleAB2}
\end{eqnarray}
Note that some NLO contribution is already involved 
in $\Sigma_{\rm eff}$ or $F_{\rm eff}$ since we have resummed
the Lagrangian with NSC vertices.

This correlator in Eq.~(\ref{eq:ppcorrelator}) is a known
result in the literature, and 
one can find how to evaluate
$\langle \mathcal{A} (U_0)\rangle_{U_0}$ and 
$\langle \mathcal{B} (U_0)\rangle_{U_0}$
in, for example, Ref.~\cite{Bernardoni:2008ei}.
In particular, the $x$ and $y$--independent constant term is known
as a special feature of the $\epsilon$ regime,
and can be used for extracting $\Sigma$.
In this work, however, we will eliminate this constant term
in the end of the calculation.
Therefore, we have to treat the second term of Eq.~(\ref{eq:ppcorrelator})
as the LO contribution, and the calculation at one order higher is needed.

\subsection{NLO calculation}
Next, let us compute the NLO contribution. 
Here and in the following, we simply neglect the
contribution to the constant part.

For the third term of Eq.~(\ref{eq:vev}),
we have
\begin{eqnarray}
\label{eq:pplosnlo}
- \langle  \langle  \left[ P(x) \right]_{12}  \left[ P(y) \right]_{21}]^{\rm LO} 
\mathcal{S}^{\rm NLO}\rangle_\xi \rangle_{U_0}
&=& - \langle  \langle [ \left[ P(x) \right]_{12}  \left[ P(y) \right]_{21}]^{\rm LO}
\mathcal{S}^{\rm NLO}_M \rangle_\xi \rangle_{U_0}
\nonumber\\&&\hspace{-1in}
=
\frac{\Sigma_{\rm eff}^2}{2F_{\rm eff}^2V}
\left\langle \mathcal{D}(U_0)\right\rangle_{U_0}
\left(-M_{12}^2\right)
\sum_{p\neq0} \frac{e^{ip(x-y)}}{(p^2)^2},
\end{eqnarray}
where $M_{12}^2 \equiv (m_1+m_2)\Sigma_{\rm eff}/F^2$, and
the dimensionless $U_0$ integral part is given by
\begin{eqnarray}
\mathcal{D}(U_0) &=& \sum_{k=0}^4\mathcal{D}^k(U_0),\\
\mathcal{D}^0(U_0) &=& [U_0]_{11}+[U_0]_{22}+[U_0^\dagger]_{11}+[U_0^\dagger]_{22},\\
\mathcal{D}^1(U_0) &=& \frac{N_f}{\mu_1+\mu_2}(2 - [U_0]_{11}[U_0]_{22}-[U_0^\dagger]_{11}[U_0^\dagger]_{22}),\\
\mathcal{D}^2(U_0) &=& \sum_{i,j} \frac{\delta_{i1}\delta_{2j}+\delta_{i2}\delta_{1j}}{2}
\left[[U_0]_{ii}\left(\frac{[U_0\mathcal{M}^\dagger U_{0}]_{jj}-m_j}{m_1+m_2}
+\frac{2N_f}{\mu_1+\mu_2}[U_0]_{jj}\right)+h.c.\right],\\
\mathcal{D}^3(U_0) &=& - \sum_{i,j}\frac{\delta_{i1}\delta_{2j}+\delta_{i2}\delta_{1j}}{N_f}
\left([U_0]_{ij}+[U_0^\dagger]_{ij}\right)
\nonumber\\&&\times
\left[\frac{[U_0\mathcal{M}^\dagger U_{0}]_{ji}+[U_0^\dagger\mathcal{M}U_{0}^\dagger]_{ji}}{m_1+m_2}
+\frac{N_f}{\mu_1+\mu_2}\left([U_0]_{ji}+[U_0^\dagger]_{ji}\right)\right],\\
\mathcal{D}^4(U_0) &=& \frac{\left([U_0]_{12}+[U_0^\dagger]_{12}\right)\left([U_0]_{21}+[U_0^\dagger]_{21}\right)}{N_f}
\left(\frac{1}{N_f}\sum_i^{N_f} \frac{m_i\left([U_0]_{ii}+[U_0^\dagger]_{ii}\right)}{m_1+m_2}
+\frac{N_f}{\mu_1+\mu_2}\right),\nonumber\\
\end{eqnarray}
where $\mu_i = m_i \Sigma_{\rm eff} V$.
Here we have given more general results than our set-up in this work:
with non-degenerate $N_f$--flavor quark masses $m_i$'s.
The degenerate results can be obtained simply taking $m_i \to m$
in the above formulas.
Note that we have neglected trivially vanishing matrix elements like
$\langle [U_0]_{ij}\rangle_{U_0} = 0$ for $i\neq j$.

The second term of Eq.~(\ref{eq:vev}) is given by
\begin{eqnarray}
\langle \langle \left[ \left[ P(x) \right]_{12} \left[ P(y) \right]_{21} \right]^{\rm NLO} \rangle_\xi \rangle_{U_0}
&=&
- \frac{\Sigma_{\rm eff}^2}{4F_{\rm eff}^4V^2}
\left\langle \mathcal{C}(U_0)\right\rangle_{U_0}
\sum_{p_1\neq0}\sum_{p_2\neq0}
\frac{e^{ip_1(x-y)}}{p_1^2}\frac{e^{ip_2(x-y)}}{p_2^2},
\end{eqnarray}
where
\begin{eqnarray}
\mathcal{C}(U_0) &=& \left(\frac{4}{N_f}-N_f\right)
\left(2 - [U_0]_{11}[U_0]_{22}-[U_0^\dagger]_{11}[U_0^\dagger]_{22}\right)
\nonumber\\&&
 +\left(1+\frac{2}{N_f^2}\right)\left([U_0]_{12}-[U_0^\dagger]_{12}\right)\left([U_0]_{21}-[U_0^\dagger]_{21}\right).
\end{eqnarray}

To summarize our results, it is useful to define
the ``massive'' propagator,
\begin{eqnarray}
\label{eq:massivepropagator}
\bar\Delta(x;M^2)
&\equiv&
\frac{1}{V} \sum_{p\neq0}
\frac{e^{ipx}}{p^2 + M^2},
\end{eqnarray}
and noting for $M\sim {\cal O}(\epsilon^2)$,
\begin{eqnarray}
\frac{1}{p^2}-M^2\frac{1}{(p^2)^2}
= \frac{1}{p^2+M^2}+{\cal O}(M^4),
\end{eqnarray}
the correlator in a simple form is obtained,
\begin{eqnarray}
\label{eq:ppresult}
\left\langle  \left[ P(x) \right]_{12} \left[ P(y) \right]_{21} \right\rangle
&=&
const. + \frac{\Sigma_{\rm eff}^2}{2F_{\rm eff}^2V}
\left\langle  \mathcal{B}(U_0)\right\rangle_{U_0}
\sum_{p\neq0} \frac{e^{ip(x-y)}}{p^2 + M_{12}^2Z_M^{\rm 2pt}}
\nonumber\\&&\hspace{1in}
-\frac{\Sigma_{\rm eff}^2}{4F_{\rm eff}^4V^2}
\left\langle \mathcal{C}(U_0) \right\rangle_{U_0}
\sum_{p_1\neq0}\sum_{p_2\neq0}
\frac{e^{ip_1(x-y)}}{p_1^2}\frac{e^{ip_2(x-y)}}{p_2^2},
\end{eqnarray}
where $const.$ denotes the constant term we have omitted, and
\begin{eqnarray}
Z_M^{\rm 2pt} &=& 1+\frac{\left\langle \mathcal{D}(U_0)
-\mathcal{B}(U_0)\right\rangle_{U_0}}{\left\langle \mathcal{B}(U_0)\right\rangle_{U_0}}.
\end{eqnarray}
The second term of Eq.~(\ref{eq:ppresult}) looks 
almost the same as the conventional massive pion
propagator in the $p$ regime.
In fact, we can smoothly obtain this by taking the $V\to \infty$ limit
where one obtains
$\left\langle \mathcal{B}(U_0)\right\rangle_{U_0}\to 4$, 
$\left\langle \mathcal{A}(U_0)\right\rangle_{U_0}=\left\langle \mathcal{C}(U_0)\right\rangle_{U_0}\to 0$,
and $Z_M^{\rm 2pt}\to 1$.

The third term of Eq.~(\ref{eq:ppresult}) is another peculiar
term in the $\epsilon$ regime, which originally comes from a 3-pion state,
consisting of one having zero momentum and two having non-zero momenta.
At this order, it looks a propagation of two massless particles.
However, these propagators should have mass corrections at higher orders,
at least, the one from the measure term in Eq.~(\ref{eq:measure}), 
We expect that it cannot reach a long-distance, compared to the single particle propagation. 
In the following analysis, we simply neglect this NLO term and
similar terms in the three-point functions.
This truncation may be numerically justified by carefully checking the plateau
of the effective mass, when we simulate lattice QCD \cite{Fukaya:2014jka}.

\subsection{Removing dominant finite volume effects in the $\epsilon$ expansion}
Now we are ready to cancel the dominant volume effects.
First, we insert spacial momentum to the operators. Namely,
we consider
\begin{eqnarray}
C^{\rm 2pt}_{PP}(t,{\bf p}) &\equiv& \langle [P(x_0;{\bf p})]_{12} [P(y_0;-{\bf p})]_{21} \rangle,
\nonumber\\
{}[P(x_0,{\bf p})]_{ij}
&\equiv&
\int d^3x\, e^{-i {\bf p\cdot x}} [P(x)]_{ij},
\end{eqnarray}
where $x_0$ is the temporal element of $x$, $t=x_0-y_0$,
and ${\bf p}=2\pi(n_x,n_y,n_z)/L$ is the 3-dimensional momentum.
Then, the unwanted constant contribution $consts.$ automatically disappears for ${\bf p}\neq {\bf 0}$.
It is also intuitively reasonable that the higher energy states having momenta
are less sensitive to the finite volume effects.
Even in the case of ${\bf p} = {\bf 0}$, it vanishes in a simple subtraction 
with respect to time: $\Delta_t [P(t,{\bf p})]_{ij}\equiv [P(t,{\bf p})]_{ij}-[P(t_{\rm ref},{\bf p})]_{ij}$
with a reference time-slice $t_{\rm ref}$.

The second step is to take a ratio of the correlators with different momenta.
For example, by shifting $y_0 \to 0$, and renaming $x_0=t$, we have
\begin{eqnarray}
\label{eq:2ptratio}
R^{\rm 2pt}(t; {\bf p}) &\equiv&  
\frac{\langle [P(t;{\bf p})]_{12} [P(0;-{\bf p})]_{21} \rangle}
{\langle \Delta_t [P(t;{\bf 0})]_{12} [P(0;{\bf 0})]_{21} \rangle}
\nonumber\\
&=& \frac{E^{\rm 2pt}({\bf 0})\sinh\left(E^{\rm 2pt}({\bf 0})T/2\right)}{E^{\rm 2pt}({\bf p})\sinh\left(E^{\rm 2pt}({\bf p})T/2\right)}
\times\frac{\cosh(E^{\rm 2pt}({\bf p})(t-T/2))}
{\cosh(E^{\rm 2pt}({\bf 0})(t-T/2))-\cosh(E^{\rm 2pt}({\bf 0})(t_{\rm ref}-T/2))},
\nonumber\\
\end{eqnarray}
where
\begin{eqnarray}
E^{\rm 2pt}({\bf p}) &\equiv& \sqrt{M_{12}^2Z_M^{\rm 2pt}+{\bf p}^2}.
\end{eqnarray}
The ratio $R^{\rm 2pt}(t; {\bf p})$ is no more dependent on 
$\langle \mathcal{A}(U_0)\rangle_{U_0}$ or $\langle \mathcal{B}(U_0)\rangle_{U_0}$.
In fact, this expression is exactly the same as the same ratio in the $p$ expansion,
except for the mass renormalization factor $Z_M^{\rm 2pt}$.
Namely, we have minimized the features of the $\epsilon$ regime
in the two-point correlator.
It is also important to note that $R^{\rm 2pt}(t; {\bf p})$ is finite
even in the limit of $E^{\rm 2pt}({\bf 0})\to 0$.

Since the above ratio $R^{\rm 2pt}(t; {\bf p})$ has no dependence on LEC's of ChPT,
it is not phenomenologically interesting.
However, it is a good test quantity for lattice QCD 
to check the validity of the above arguments.
Recently, JLQCD collaboration \cite{Fukaya:2014jka} 
compared the ratio $R^{\rm 2pt}(t; {\bf p})$ to 
the numerical data in the both cases with $M_{12}\sqrt{Z_M^{\rm 2pt}}=0$ and 100 MeV
and found a fairly good agreement, which suggests that 
the NLO corrections in $\sqrt{Z_M^{\rm 2pt}}$ and the third term 
of Eq.~(\ref{eq:ppresult}) we have neglected are actually small.

Since the $x$-independence of the pion zero-mode and its dominance in
the finite volume effects are universal and true in any correlation functions 
at any sizes of the volume,
we expect wide applications of our method.
Namely, inserting momenta to the correlators and taking a ratios of them
generally makes a less sensitive quantity to the volume than the original ones.
We will see this is true for the three-point functions in the next section.

\section{Three-point function}
\label{sec:threepoint}

In this section, we calculate our main target, 
the pseudoscalar-vector-pseudoscalar three-point function 
in a finite volume in the $\epsilon$ expansion of ChPT, 
which is relevant for extracting the vector pion form factor. 
However, we should note that the 
pion form factor itself is not a quantity 
described within ChPT alone.
In numerical studies \cite{Aoki:2009qn, Kaneko:2010ru}
it is known that the vector meson largely contributes to the results,
which cannot be explained by ChPT.
Even in such a case, we still expect that the correction from the finite volume
can be treated within ChPT, as the heavier hadrons, including the vector mesons,
do not propagate very long.
Therefore, in this section, we compute the finite volume effects on
the three-point function within the $\epsilon$ expansion of ChPT.
Once the main part of finite volume effects are removed, 
the remaining pion form factor should
include the physics beyond ChPT.

\subsection{Three-point functions and form factors}

First, we briefly review how the three-point functions
are related to the pion form factors.
Our main target in this work is the vector form factor,
defined by
\begin{eqnarray}
\langle \pi^a(p_2) | V^b_{\mu}(x) | \pi^c(p_1)\rangle 
&=&
i\epsilon^{abc}(p_1+p_2)_\mu F_V(t),
\end{eqnarray}
where $| \pi^a(p)\rangle$ denotes the on-shell pion state
with momentum $p$, $V^b_{\mu}(x)$ is 
the coefficient of an $SU(2)$ generator $\tau^b$ 
in the vector operator, and $t=(p_1-p_2)^2$.

For lattice QCD calculations, it is convenient 
to take $b=3$ component
\begin{eqnarray}
V^3_{\mu}(x) &=& \frac{1}{2}\bar{u}\gamma_\mu u(x)-\frac{1}{2}\bar{d}\gamma_\mu d(x).
\end{eqnarray}
Using a conventional notation
\begin{eqnarray}
 | \pi^1(p)\rangle &=& \frac{| \pi^+(p)\rangle+| \pi^-(p)\rangle}{\sqrt{2}},\;\;\;\;
  | \pi^2(p)\rangle = \frac{| \pi^+(p)\rangle-| \pi^-(p)\rangle}{\sqrt{2}i},
\end{eqnarray}
where $| \pi^\pm (p)\rangle$ denotes the charged pion state,
and iso-spin symmetry (we assume $m_u=m_d=m$),
\begin{eqnarray}
\langle \pi^+(p_2) | V^3_{\mu}(x) | \pi^+(p_1)\rangle 
&=&
- \langle \pi^-(p_2) | V^3_{\mu}(x) | \pi^-(p_1)\rangle, 
\end{eqnarray}
as well as the electric charge conservation,
\begin{eqnarray}
\langle \pi^+(p_2) | V^3_{\mu}(x) | \pi^-(p_1)\rangle 
&=&
\langle \pi^-(p_2) | V^3_{\mu}(x) | \pi^+(p_1)\rangle=0, 
\end{eqnarray}
one obtains a simpler formula,
\begin{eqnarray}
\langle \pi^+(p_2) | V^3_{\mu}(x) | \pi^+(p_1)\rangle 
&=& (p_1+p_2)_\mu F_V(t).
\end{eqnarray}

It is also important to note for the isospin zero current,
\begin{eqnarray}
V^0_{\mu}(x) &=& \bar{u}\gamma_\mu u(x)+\bar{d}\gamma_\mu d(x),
\end{eqnarray}
that its form factor is zero:
\begin{eqnarray}
\langle \pi^a(p_2) | V^0_{\mu}(x) | \pi^b(p_1)\rangle  =0 \;\;\;\mbox{for any $a,b$},
\end{eqnarray}
since the pions have zero Baryon charge.
In ChPT, this situation is more directly shown by
$V^0_{\mu}(x)={\rm Tr}\;V_\mu(x)=0$ in Eq.~(\ref{eq:V}).
Namely, there exists no corresponding current within ChPT.
Therefore, for the electro-magnetic current defined by
\begin{eqnarray}
J_\mu^{EM} &\equiv& V_\mu^3(x)+\frac{1}{6}V_\mu^0(x)
=\frac{2}{3}\bar{u}\gamma_\mu u(x)-\frac{1}{3}\bar{d}\gamma_\mu d(x),
\end{eqnarray}
one can show an identity,
\begin{eqnarray}
\langle \pi^+(p_2) | V^3_{\mu}(x) | \pi^+(p_1)\rangle 
&=& \langle \pi^+(p_2) | J^{EM}_{\mu}(x) | \pi^+(p_1)\rangle.
\end{eqnarray}
Namely, we don't have to distinguish the vector form factor
from the electro-magnetic form factor of the pions.

In the literature, the finite volume correction on 
the hadronic matrix elements
is often computed by just replacing the quantum  
loop momentum integrals by a discrete summation.
However, in such a calculation, one assumes
that one can apply the same LSZ reduction formula as in the $V\to \infty$ limit, 
to relate the form factor to the three-point function,
\begin{eqnarray}
\int d^4 x e^{ip_2 x} \int d^4 z e^{-ip_1 z}
\langle \left[P(x)\right]_{12} V^3_{\mu}(y) \left[P(z)\right]_{21}\rangle&&
\nonumber\\&&\hspace{-2in} 
=\frac{\langle 0|\left[P(0)\right]_{12}|\pi^+(p_2)\rangle\langle \pi^+(p_1)|\left[P(0)\right]_{21}|0\rangle}{(p_1^2+m_\pi^2)(p_2^2+m_\pi^2)}
\langle \pi^+(p_2) | V^3_{\mu}(y) | \pi^+(p_1)\rangle.
\end{eqnarray}
In a finite volume (simulated on the lattice),
this relation is non-trivial, and one may overlook
finite volume corrections to the reduction formula itself.
In this work, we work on the finite volume correction within ChPT to
\begin{eqnarray}
\langle \left[P(x)\right]_{12} [V_{\mu}(y)]_{ii}\left[P(z)\right]_{21}\rangle,
\end{eqnarray}
with a general flavor index $i$.
We will soon see that $\langle \left[P(x)\right]_{12} [V_{\mu}(y)]_{ii}\left[P(z)\right]_{21}\rangle
=(\delta_{i1}-\delta_{i2})\langle \left[P(x)\right]_{12} V^3_{\mu}(y)\left[P(z)\right]_{21}\rangle$.
We then perform its Fourier transformation with non-zero momenta,
and show how to disentangle the pion form factor from the
correlators.

\subsection{LO contribution}
In the following, we assume $x_0>y_0>z_0$, and denote $t=x_0 - y_0$, $t^\prime = y_0-z_0$.
We further assume that $t$, $t^\prime$, $t+t^\prime < T/2$ to suppress the
effect of modes wrapping around our periodic lattice.
It is straightforward to compute the LO contribution to the three-point function 
in the same way as the two-point function,
\begin{eqnarray}
\langle \left[ P(x) \right]_{12} \left[ V_\mu(y) \right]_{ii} \left[ P(z) \right]_{21}\rangle
&=& 
(\delta_{i2}-\delta_{i1})
\left(\frac{Z^{V2}}{(Z^\xi)^2}\right)\frac{i\Sigma_{\rm eff}^2}{4F_{\rm eff}^2V^2}
\left\langle \mathcal{E}(U_0)\right\rangle_{U_0}
\nonumber\\&&\times
\sum_{p_1\neq0}\sum_{p_2\neq0}
\frac{-ip_1^\mu -ip_2^\mu}{p_1^2\, p_2^2}
e^{ip_1(x-y)} e^{ip_2(y-z)},
\end{eqnarray}
where 
\begin{eqnarray}
\mathcal{E}(U_0) &=& 
\left(2+2[U_0]_{11}[U_0]_{22}+2[U_0^\dagger]_{11}[U_0^\dagger]_{22}
\right.\nonumber\\&&\left.
+[U_0]_{11}[U_0^\dagger]_{11}+[U_0]_{22}[U_0^\dagger]_{22}
-[U_0]_{12}[U_0^\dagger]_{21}-[U_0]_{21}[U_0^\dagger]_{12}\right).
\end{eqnarray}
Here, we have neglected the $t$ and $t^\prime$ independent terms
since we will automatically cancel them in the end of our computation.

We have also neglected diagrams where $\xi$'s are connected 
in unusual orders, like $x$--$z$--$y$ or $z$--$x$--$y$, expecting
the long propagation between $x$ and $z$ to be 
exponentially suppressed.
This expectation is not true for the zero-momentum contribution at LO.
However, as mentioned in the previous section,
it is reasonable to expect that the NLO corrections give
a ``mass'' to the correlators and make long-range correlation 
suppressed compared to the main result.
One should be able to numerically check this expectation,
since if the neglected contribution is big, 
it should be detected as unexpected $|x-z|$ dependence.

\subsection{NLO contribution}

Next, let us calculate the NLO corrections to the three-point function. 
As seen in the two-point function,
the contribution from $\mathcal{S}^{\rm NLO}_M$ can be
encoded as the mass corrections: 
together with the LO contribution, one can express it as
\begin{eqnarray}
\label{eq:pvpSM}
\langle\langle \left[ P(x) \right]_{12} \left[ V_\mu(y) \right]_{ii} \left[ P(z) \right]_{21} ]^{\rm LO}(1-\mathcal{S}^{\rm NLO}_M)\rangle_\xi \rangle_{U_0}
\nonumber\\&&\hspace{-3.2in}
=(\delta_{i2}-\delta_{i1})
\frac{i\Sigma_{\rm eff}^2}{4F_{\rm eff}^2V^2}
\left\langle \mathcal{E}(U_0)\right\rangle_{U_0}
\sum_{p_1\neq0}\sum_{p_2\neq0}
\frac{-ip_1^\mu -ip_2^\mu}
{(p_1^2 + M_{12}^2 Z_M^{\rm 3pt}) (p_2^2 + M_{12}^2 Z_M^{\rm 3pt})}
e^{ip_1(x-y)}e^{ip_2(y-z)},
\nonumber\\
\end{eqnarray}
where
\begin{eqnarray}
Z_M^{\rm 3pt} &=&  1+\frac{N_f}{M_{12}^2F^2 V}
+\frac{\langle\mathcal{G}(U_0)+\mathcal{H}(U_0)\rangle_{U_0}}{\langle \mathcal{E}(U_0)\rangle_{U_0}},
\end{eqnarray}
\begin{eqnarray}
\mathcal{G}(U_0)
&\equiv&
 \frac{1}{4}\left[
\left\{([U_0]_{22}+[U_0^\dagger]_{22}-2)
(2+[U_0]_{11}[U_0]_{22}+[U_0^\dagger]_{11}[U_0^\dagger]_{22})
\right.\right.\nonumber\\&&\left.\left.
\hspace{0.5in}
+8([U_0]_{22}+[U_0^\dagger]_{22})
\right.\right.\nonumber\\&&\left.\left.
\hspace{0.5in}
-6[U_0]_{11}[U_0]_{22}-6[U_0^\dagger]_{11}[U_0^\dagger]_{22}
-4[U_0]_{22}[U_0^\dagger]_{22}
\right.\right.\nonumber\\&&\left.\left.
\hspace{0.5in}
-([U_0]_{22}+[U_0^\dagger]_{22}-4)
([U_0]_{12}[U_0^\dagger]_{21}+[U_0^\dagger]_{12}[U_0]_{21})
\right.\right.\nonumber\\&&\left.\left.
\hspace{0.5in}
-([U_0]_{12}[U_0]_{21}[U_0]_{22}
+[U_0^\dagger]_{12}[U_0^\dagger]_{21}[U_0^\dagger]_{22})
\right.
\right.\nonumber\\&&\left.\left.
\hspace{0.5in}+
2([U_0]_{11}+[U_0^\dagger]_{11}-2)(1+[U_0]_{22}[U_0^\dagger]_{22})
\right\}\right.
\nonumber\\&&\left.
\hspace{0.5in}+
([U_0]_{11}+[U_0^\dagger]_{22})([U_0\mathcal{M}^\dagger U_0]_{22}/m-1)
\right.\nonumber\\&&\left.
\hspace{0.5in}
+([U_0^\dagger]_{11}+[U_0]_{22})([U_0^\dagger\mathcal{M}U_0^\dagger]_{22}/m-1)
\right.\nonumber\\&&\left.
\hspace{0.5in}+
2 [U_0]_{22} ([U_0\mathcal{M}^\dagger U_0]_{11}/m-1)
+2 [U_0^\dagger]_{22} ([U_0^\dagger\mathcal{M}U_0^\dagger]_{11}/m-1)
\right.\nonumber\\&&\left.
\hspace{0.5in}
\textcolor{black}{
-([U_0]_{12}[U_0^\dagger]_{21}[U_0^\dagger]_{22}
+[U_0^\dagger]_{12}[U_0]_{21}[U_0]_{22})
}
\right.\nonumber\\&&\left.
\hspace{0.5in}
\textcolor{black}{
-([U_0^\dagger]_{21} [U_0\mathcal{M}^\dagger U_0]_{12}/m
+[U_0]_{21} [U_0^\dagger\mathcal{M}U_0^\dagger]_{12}/m)
}
\right],
\\
\mathcal{H}(U_0) &\equiv&
-\frac{1}{2N_f}\left[
([U_0]_{12}+[U_0^\dagger]_{12})
([U_0\mathcal{M}^\dagger U_0]_{21}/m+[U_0^\dagger \mathcal{M} U_0^\dagger]_{21}/m)
\right].
\end{eqnarray}


For the correction in the operators, we have a contribution from the $L_9$ term:
\begin{eqnarray}
\label{eq:L9term}
\langle\langle [ \left[ P(x) \right]_{12} \left[ V_\mu(y) \right]_{ii} 
\left[ P(x) \right]_{21}]^{L_9}\rangle_\xi \rangle_{U_0}
&=&
\nonumber\\&&\hspace{-1.6in}
(\delta_{i2}-\delta_{i1})\frac{i\Sigma_{\rm eff}^2}{4F_{\rm eff}^2V^2}
\left\langle \mathcal{E}(U_0)\right\rangle_{U_0}
\left( -\frac{2L_9}{F_{\rm eff}^2} \right)
\nonumber\\&&\hspace{-1.6in}
\times
\sum_{p_1\neq0}\sum_{p_2\neq0}
\frac{i\left[p_2\cdot(p_1-p_2)\right](p_1)_\mu
- i\left[p_1\cdot(p_1-p_2)\right](p_2)_\mu}
{p_1^2\, p_2^2}
e^{ip_1(x-y)} e^{ip_2(y-z)},
\nonumber\\
\end{eqnarray}

The correction from $\mathcal{S}_K^{\rm NLO}$ term is obtained as
\begin{eqnarray}
\label{eq:NLOSK}
-\langle\langle \left[ P(x) \right]_{12} \left[ V_\mu(y) \right]_{ii} 
\left[ P(z) \right]_{21}]^{\rm LO}\mathcal{S}^{\rm NLO}_K\rangle_\xi \rangle_{U_0}
&=& \nonumber\\&&\hspace{-2.0in}
(\delta_{i2}-\delta_{i1})
\frac{i\Sigma_{\rm eff}^2}{4F_{\rm eff}^2V^2}
\left\langle\mathcal{E}(U_0)\right\rangle_{U_0}
\nonumber\\&&\hspace{-2.0in}
\times
\left(-\frac{N_f}{2F_{\rm eff}^2} \right)
\sum_{p_1\neq0} \sum_{p_2\neq0}
\frac{-i(p_1 + p_2)^\nu I_{\mu\nu}(-p_1^0 + p_2^0,-{\bf p}_1 + {\bf p}_2)}
{p_1^2\, p_2^2}
e^{ip_1(x-y)}e^{ip_2(y-z)},
\nonumber\\
\end{eqnarray}
where
\begin{eqnarray}
I_{\mu\nu}(q_0,{\bf q})
&\equiv&
\frac{1}{V} \sum_{p\neq0,q}
\frac{p^\mu (q^\nu - 2p^\nu)}{p^2 (q-p)^2}
\;\;\; (q^2=q_0^2+{\bf q}^2).
\end{eqnarray}


Now let us summarize all the above results for the $\mu=0$ case,
inserting momenta ${\bf p}_f$ and ${\bf p}_i$.
Using the notations $t=x_0-y_0$, $t^\prime = y_0-z_0$, 
$E^{\rm 3pt}({\bf p}) \equiv \sqrt{M_{12}^2 Z_M^{\rm 3pt}+{\bf p}^2}$,
and
\begin{eqnarray}
c({\bf p}, t) &=& \frac{\cosh[E^{\rm 3pt}({\bf p})(t-T/2)]}{2E^{\rm 3pt}({\bf p})\sinh[E^{\rm 3pt}({\bf p})T/2]},\;\;\;\;\;
s({\bf p}, t) = \frac{\sinh[E^{\rm 3pt}({\bf p})(t-T/2)]}{2E^{\rm 3pt}({\bf p})\sinh[E^{\rm 3pt}({\bf p})T/2]},
\end{eqnarray}
one can express the result as
\begin{eqnarray}
\label{eq:PVPgeneral}
C^{PV_0P}(t, t^\prime;  {\bf p}_f, {\bf p}_i) &\equiv&
\langle \left[P(x_0, -{\bf p}_f)\right]_{12} V^3_0(y_0, {\bf q}) \left[P(z_0, {\bf p}_i)\right]_{21}\rangle 
\hspace{-2in}
\nonumber\\&=& 
-\frac{L^3\Sigma_{\rm eff}^2}{4F_{\rm eff}^2}
\left\langle \mathcal{E}(U_0)\right\rangle_{U_0}
\delta^{(3)}_{{\bf q},{\bf p}_f-{\bf p}_i}Z_{k} F_V(q_0,{\bf q})
\nonumber\\&&
\times \left[iE^{\rm 3pt}({\bf p}_i)c({\bf p}_f, t)s({\bf p}_i, t^\prime)
+iE^{\rm 3pt}({\bf p}_f)s({\bf p}_f, t)c({\bf p}_i, t^\prime)
\right].
\end{eqnarray}
Here, as mentioned in the above calculations, 
we have omitted the two-pion-like propagations,
and the $x_0-z_0=t+t^\prime$ dependent long-distance 
correlators, as they are expected to be exponentially small.

The vector form factor $F_V (q_0,{\bf q})$ is given by
\begin{eqnarray}
\label{eq:FVdef}
F_V(q_0,{\bf q}) &=& \frac{Z^{V2}}{(Z^\xi)^2}- \frac{2 L_9}{F_{\rm eff}^2}q^2
-\frac{N_f}{2F_{\rm eff}^2}\left(l(q_0,{\bf q}) 
 - l_{00}\right),
\end{eqnarray}
where $l(q_0, {\bf q})$ is a part of $I_{0\nu}(q_0,{\bf q})$
which is proportional to $\delta_{0\nu}$.
Another part proportional to $q_0q_\nu$ cannot 
contribute since it is contracted with a perpendicular
vector $\bar{q}^\nu$ to $q_\mu$. Namely, $l(q_0, {\bf q})$ is given by
\begin{eqnarray}
l (q_0, {\bf q}) &=&
I_{0\nu}(q_0,{\bf q}) \bar{q}^\nu / \bar{q}_0.
\end{eqnarray}
More details are discussed in Appendix~\ref{app:Iintegral}.

Note in the above formula, the (finite) renormalization factor
\begin{eqnarray}
Z_k &=& 1-\frac{N_f}{2F_{\rm eff}^2} l_{00}, \;\;\;\;\;
l_{00}
\equiv
-\frac{1}{4\pi^2}
\sum_{b \neq0}
\frac{1}{|b_\mu|^2} \left(1 - \frac{2 (b_0)^2}{|b_\mu|^2} \right),
\end{eqnarray}
where the summation is taken over the 
vector $b_\mu = (n_0 T, n_1L, n_2L,n_3L)$ with integers $n_\mu$,
is introduced so that $F_V(0,{\bf 0})=1$ is maintained
even in a finite volume. 
Therefore, the finite volume effects contained in 
$F_V(q_0,{\bf q})$ are only those which come
from the non-zero modes, 
vanish in the $q_\mu\to 0$ limit, and are thus expected
to be perturbatively small.
We will discuss the details of the remaining finite volume effects in the next section.

Finally, let us discuss the renormalization of the
above formula Eq.~(\ref{eq:FVdef}). Since the finite volume
effects are free from UV divergences,
it is sufficient to consider the $V\to \infty$ limit
of $F_V(q_0,{\bf q})$.
It is not difficult to see that the quadratic
divergence in $Z^{V2}/(Z^\xi)^2$ is precisely
canceled by that in $l(q_0, {\bf q})$.
Therefore, we only need to renormalize the 
logarithmic divergence of $l(q_0, {\bf q})$
by the re-definition of $L_9$.

Employing the dimensional regularization, we can easily evaluate
its logarithmic divergence as
\begin{eqnarray}
\label{eq:Imunuinfty}
\lim_{V\to\infty} l(q_0,{\bf q})
&=& \frac{\bar{q}^\nu}{\bar{q}_0}
\int \frac{d^dp}{(2\pi)^d} \frac{-2p_0 p_\nu}{p^2 (p-q)^2}
\nonumber\\
&=&
\frac{1}{16\pi^2}
\left\{ \frac{q^2}{6} \left( \frac{2}{\epsilon} +1-\gamma_{E}
+\ln4\pi - \ln\mu_{sub}^2\right)
-\frac{q^2}{6} \ln\frac{q^2}{\mu_{sub}^2} +\frac{5}{18}q^2\right\},
\end{eqnarray}
where $\epsilon=4-d$, $\gamma_E=0.57721\cdots$ is the Euler's constant,
and $\mu_{sub}$ denotes the subtraction scale. 
This divergence can be absorbed in the renormalization of $L_9$:
\begin{eqnarray}
L_9^r (\mu_{sub})&\equiv&
L_9 - \frac{N_f}{12}\times \frac{1}{16\pi^2}
\left(-\frac{1}{\epsilon}-\frac{1}{2}(-\gamma_E +\ln 4\pi+1
-\ln \mu_{sub}^2)\right),
\end{eqnarray}
and one obtains the infinite volume limit for the vector form factor,
\begin{eqnarray}
F_V^\infty(q_0,{\bf q}) &=& 
1 - \frac{2 L^r_9(\mu_{sub})}{F_{\rm eff}^2}q^2-\frac{N_f}{2F_{\rm eff}^2}
\frac{1}{16\pi^2}
\left[
-\frac{1}{6}q^2 \ln \frac{q^2}{\mu_{sub}^2}+\frac{5}{18}q^2
\right],
\end{eqnarray}
which agrees with the known (massless limit of) 
result within ChPT.
Note that we cannot expect $F_V^\infty(q_0,{\bf q})$ to
describe the lattice data well, since the physics beyond ChPT is omitted
in the ChPT expression.
However, we can still expect that the finite volume correction :
$F_V(q_0,{\bf q})-F_V^\infty(q_0,{\bf q})$
is well described within ChPT, which will be discussed in the next section.

\section{Extraction of the vector form factor of pion}
\label{sec:extraction}

In this section, we show how to eliminate the leading
zero-momentum pion mode's contribution from the
correlator, and how to extract the vector form
factor of pions. There still remain
finite volume effects from non-zero modes
but they are sub-leading contributions.
From the one-loop calculation of the
non-zero momentum modes, we numerically 
estimate this remaining effect, and 
show they are actually a small perturbation.

\subsection{Removing dominant finite volume effects from the pion zero mode}

In the previous section, we have neglected the $t$--independent or 
$t^\prime$--independent terms in our calculation.
In the final form Eq.~(\ref{eq:PVPgeneral}), 
if both of ${\bf p}_i$ and ${\bf p}_f$ are non-zero, 
these terms are automatically dropped.
However, if these momenta are zero, 
we have to take subtraction of the correlators 
at different time-slices,
$\Delta_t f(t)\equiv f(t)-f(t_{\rm ref})$,
$\Delta_{t^\prime} f(t^\prime)\equiv f(t^\prime)-f(t^\prime_{\rm ref})$,
with $t_{\rm ref}$ and $t_{\rm ref}^\prime$, respectively.
Similar procedure was already shown in the two-point correlators.
To keep $t_{\rm ref}+t^\prime_{\rm ref} < T/2$
and $t, t^\prime < t_{\rm ref}$, which are the conditions
to suppress the contribution from pions wrapping around the periodic space-time,
$t_{\rm ref}=t_{\rm ref}^\prime\sim T/4$ would be optimal.
In the following, we take $t^\prime_{\rm ref}=t_{\rm ref}$, for simplicity.

With the above time-slice subtraction in mind, and noting $F_V(0,{\bf 0})=1$ 
the following ratios are useful for 
extracting the vector pion form factor:
\begin{eqnarray}
R_1(t, t^\prime;  {\bf p}_f, {\bf p}_i) &\equiv&
\frac{C^{PV_0P}(t, t^\prime;  {\bf p}_f, {\bf p}_i)}{\Delta_t \Delta_{t^\prime}C^{PV_0P}(t, t^\prime;  {\bf 0}, {\bf 0})}
\nonumber\\
&=& F_V(q_0,{\bf q}) \times \frac{E^{\rm 3pt}({\bf p}_i)c({\bf p}_f, t)s({\bf p}_i, t^\prime)
+E^{\rm 3pt}({\bf p}_f)s({\bf p}_f, t)c({\bf p}_i, t^\prime)}{E^{\rm 3pt}({\bf 0})\Delta_t c({\bf 0}, t)\Delta_{t^\prime} s({\bf 0}, t^\prime)
+E^{\rm 3pt}({\bf 0})\Delta_t s({\bf 0}, t)\Delta_{t^\prime} c({\bf 0}, t^\prime)},
\nonumber\\
R_2(t, t^\prime;  {\bf 0}, {\bf p}_i) &\equiv&
\frac{\Delta_t C^{PV_0P}(t, t^\prime;  {\bf 0}, {\bf p}_i)}{\Delta_t \Delta_{t^\prime}C^{PV_0P}(t, t^\prime;  {\bf 0}, {\bf 0})}
\nonumber\\
&=& F_V(q_0,{\bf q}) \times \frac{E^{\rm 3pt}({\bf p}_i)\Delta_t c({\bf 0}, t)s({\bf p}_i, t^\prime)
+E^{\rm 3pt}({\bf 0})\Delta_t s({\bf 0}, t)c({\bf p}_i, t^\prime)}{E^{\rm 3pt}({\bf 0})\Delta_t c({\bf 0}, t)\Delta_{t^\prime} s({\bf 0}, t^\prime)
+E^{\rm 3pt}({\bf 0})\Delta_t s({\bf 0}, t)\Delta_{t^\prime} c({\bf 0}, t^\prime)}.
\nonumber\\
\end{eqnarray}
Note here that the $t$ and $t^\prime$ dependences are
uniquely determined once $M_{12}\sqrt{Z_M^{\rm 3pt}}$ is given.
Therefore, $F_V(q_0,{\bf q})$ can be extracted by performing a one-parameter fit
at a long-distance, taking $M_{12}\sqrt{Z_M^{\rm 3pt}}$ as a free parameter.
 
\begin{figure}[htbp]
\vspace{-20pt}
   \begin{center}
     \includegraphics[bb=0 0 260 162,width=10cm]{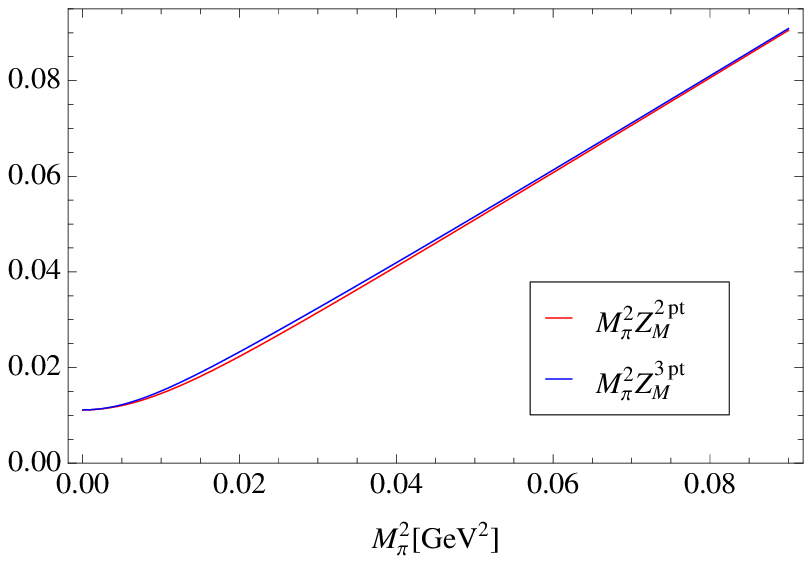}  
     \vspace{-10pt}
  \end{center}
\vspace{0pt}
   \begin{center}
     \includegraphics[bb=0 0 260 162,width=10cm]{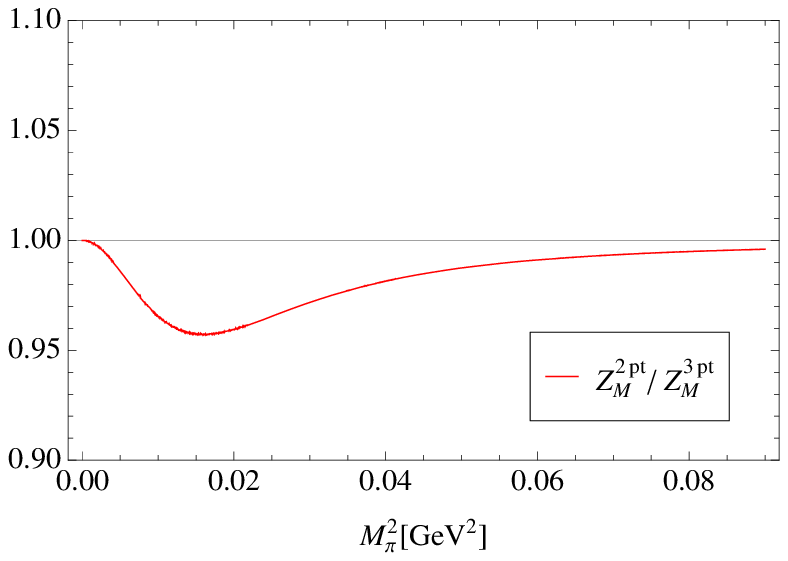}  
     \vspace{-10pt}
     \caption{Numerical estimates for the {\it pion mass} squared 
       $M_\pi^2 Z_M^{2pt}$ and $M_\pi^2 Z_M^{3pt}$ (top panel) and their ratio $Z_M^{2pt} / Z_M^{3pt}$ (bottom).
       Here, we use $L=T/2$=2fm, and $F_{\rm eff}=92.2$ MeV as inputs. 
}
     \label{fig:Z2pt_Z3pt}
  \end{center}
\end{figure}

In the numerical lattice analysis, one could also try taking further ratios with two-point functions.
Namely, 
\begin{eqnarray}
R_1^\prime(t, t^\prime;  {\bf p}_f, {\bf p}_i) &\equiv&
\frac{C^{PV_0P}(t, t^\prime;  {\bf p}_f, {\bf p}_i)}{\Delta_t \Delta_{t^\prime}C^{PV_0P}(t, t^\prime;  {\bf 0}, {\bf 0})}
\nonumber\\&&
\times \left(
\frac{-\Delta_t C^{\rm 2pt}_{PP}(t,{\bf 0})\Delta_{t^\prime} 
    \partial_{t^\prime} C^{{\rm 2pt}}_{PP}( t^\prime ,{\bf 0})
    -\Delta_t \partial_t C^{\rm 2pt}_{PP}(t,{\bf 0})\Delta_{t^\prime}
    C^{{\rm 2pt}}_{PP}( t^\prime ,{\bf 0})}{
(E^{\rm 2pt}({\bf p}_i)+E^{\rm 2pt}({\bf p}_f))C^{\rm 2pt}_{PP}(t,{\bf p}_i)C^{\rm 2pt}_{PP}(t^\prime,{\bf p}_f)
}\right),
\nonumber\\
R_2^\prime(t, t^\prime;  {\bf 0}, {\bf p}_i) &\equiv&
\frac{\Delta_t C^{PV_0P}(t, t^\prime;  {\bf 0}, {\bf p}_i)}{\Delta_t \Delta_{t^\prime}C^{PV_0P}(t, t^\prime;  {\bf 0}, {\bf 0})}\nonumber\\&&
\times \left(
\frac{-\Delta_t C^{\rm 2pt}_{PP}(t,{\bf 0})\Delta_{t^\prime} 
    \partial_{t^\prime} C^{{\rm 2pt}}_{PP}( t^\prime ,{\bf 0})
    -\Delta_t \partial_t C^{\rm 2pt}_{PP}(t,{\bf 0})\Delta_{t^\prime}
    C^{{\rm 2pt}}_{PP}( t^\prime ,{\bf 0})}{C^{\rm 2pt}_{PP}( t^\prime,{\bf p}_i)
    \left[
      -\Delta_{t} \partial_{t} C^{{\rm 2pt}}_{PP}( t ,{\bf 0})
      +E({\bf p}_i)\Delta_{t} C^{{\rm 2pt}}_{PP}( t ,{\bf 0})
    \right]
}\right).
\nonumber\\
\end{eqnarray}
Note that $E^{\rm 2pt}({\bf p})=E^{\rm 3pt}({\bf p})$ at LO.
At NLO, their expressions are different, 
reflecting the different zero-mode integrals.
However, they are numerically very similar to each other
with reasonable set-ups of the lattice simulation parameters.
In particular, they share the exactly same chiral limit,
and the infinite volume limit as seen in Figure \ref{fig:Z2pt_Z3pt}.
Therefore, these ratios $R_1^\prime(t, t^\prime;  {\bf p}_f, {\bf p}_i)$,
and $R_2^\prime(t, t^\prime;  {\bf 0}, {\bf p}_i)$ should cancel the
$t$ and $t^\prime$ dependences, and
directly give the values of $F_V(q_0,{\bf q})$. 

JLQCD collaboration \cite{Fukaya:2014jka} has employed the latter ratios
and found a good plateau for it, extracting
a pion charge radius, which is consistent with the experiment. 

It should be noted that except for $Z_M^{\rm 3pt}$,
which is essentially irrelevant in both of the above ratios,
we don't need any zero-mode integrals which could 
have been a complicated combination of Bessel functions.
The remaining finite volume effect in $F_V(q_0,{\bf q})$
is a perturbative correction from the non-zero modes only
and thus, is expected to be small as shown in the next subsection.

\subsection{Remaining Finite Volume Effects from non-zero modes}

After removing the dominant finite volume effect from
the zero-mode, what remains in $F_V(q_0,{\bf q})$ is
the effect of the non-zero momentum modes, which 
is expected to be perturbatively small.
In this subsection, we compute this non-zero-momentum 
effect to the pion 1-loop and numerically confirm
this expectation.

To this end, all we need to evaluate is 
\begin{eqnarray}
\label{eq:Imunu}
I_{\mu\nu} (q_0,{\bf q}) &=&
\frac{1}{V}
\sum_{p\neq 0, q}\frac{-2p_\mu p_\nu}{p^2(p-q)^2}.
\end{eqnarray}
Here and in the following, we ignore the terms
proportional to $q_\nu$, since they are
always contracted with a perpendicular 4-momentum vector to $q_\mu$,
and thus do not contribute to the final result.

It is not difficult to decompose it as
\begin{eqnarray}
I_{\mu\nu} (q_0,{\bf q}) &=&
\sum_{b_\mu = n_\mu L_\mu} 
I^b_{\mu\nu} (q_0,{\bf q}),
\end{eqnarray}
where 
\begin{eqnarray}
I^b_{\mu\nu} (q_0,{\bf q}) \equiv 
\int \frac{d^4p}{(2\pi)^4} e^{ipb} 
\frac{-2p_\mu p_\nu}{p^2(p-q)^2}.
\end{eqnarray}
Note that $I^{b=0}(q_0,{\bf q})$ is the infinite volume
limit of $I_{\mu\nu} (q_0,{\bf q})$ and thus,
the finite volume correction is given by
\begin{eqnarray}
\label{eq:DeltaI}
\Delta I_{\mu\nu} (q_0,{\bf q}) &=&
\sum_{b\neq 0} I^b_{\mu\nu} (q_0,{\bf q}).
\end{eqnarray}

In the standard manner, each contribution $I^b_{\mu\nu} (q_0,{\bf q})$
can be computed as
\begin{eqnarray}
\label{eq:I^b}
I^b_{\mu\nu} (q_0,{\bf q})
&=& 2\frac{\partial}{\partial b^\mu}\frac{\partial}{\partial b^\nu}
\int_0^1 dx e^{ix bq} \int \frac{d^4p}{(2\pi)^4} \frac{e^{ipb}}{(p^2+\Delta)^2}
\nonumber\\
&=& -\frac{1}{4\pi^2}
\int_0^1 dx e^{ix bq} \left[
\frac{\delta_{\mu\nu}}{|b_\mu|}
\sqrt{\Delta}K_1(\sqrt{\Delta}|b_\mu|)
-\frac{b_\mu b_\nu}{|b_\mu|^2}\Delta K_2(\sqrt{\Delta}|b_\mu|)
\right],
\nonumber\\
\end{eqnarray}
where $\Delta=x(1-x)q^2$, and $K_i(z)$ denotes the $i$-th modified
Bessel function.
Here, we have neglected a term proportional to $q_\mu b_\nu$, since
that term is proportional to $q_\nu$ after the summation over $b_\nu$.

When $b_0 =0$, it is straight forward to 
numerically evaluate the above form.
However, when $b_0\neq 0$, we need a special 
care because we need to analytically continuate
the results with respect to $q_0$.
Here we simplify the situation using an inequality
\begin{eqnarray}
\left| \int_0^1 dx e^{i\alpha} f(x)\right| < \left|\int_0^1 dx |e^{i\alpha}| f(x)\right| = \left|\int_0^1 dx  f(x)\right|,
\end{eqnarray}
in Eq.~(\ref{eq:I^b}).
Namely we neglect the oscillating factor $\exp(ixb_0 q_0)$.
Then the analytic continuation of $q_0$ has no subtlety
since the Bessel functions are all vanishing 
in the limit $|q_0| \to \infty$ with any complex phase.
Note here that the real part $\sqrt{\Delta}$ is always positive.
We do not think this {\it over-estimation}
affects the result very much, 
since the temporal direction is usually larger than
the spacial direction by a factor of 2 or 3, 
and therefore, the contribution from $b_0\neq 0$ is much smaller from the beginning.

Taking $\mu=0$ direction the finite volume correction to $F_V(q_0,{\bf q})$
can be computed as
\begin{eqnarray}
\label{eq:DeltaF}
\Delta F_V(q_0,{\bf q}) &\equiv & F_V(q_0,{\bf q})-F^\infty_V(q_0,{\bf q})
\nonumber\\
&=&
-\frac{N_f}{2F_{\rm eff}^2} \left( \Delta l (q_0,{\bf q}) -l_{00} \right),
\end{eqnarray}
where 
\begin{eqnarray}
\label{eq:I^bevaluation}
\Delta l (q_0,{\bf q}) &=& 
-\frac{1}{4\pi^2}\sum_{b_\mu}
\int_0^1 dx e^{ix {\bf b}\cdot {\bf q}}\left[
\frac{\sqrt{\Delta}}{|b_\mu|}K_1(\sqrt{\Delta}|b_\mu|)
-\frac{b_0^2}{|b_\mu|^2}\Delta K_2(\sqrt{\Delta}|b_\mu|)
\right].
\end{eqnarray}
Note that $\Delta l (0,{\bf 0})=l_{00}$.

\begin{figure}[htbp]
\vspace{-20pt}
   \begin{center}
     \includegraphics[bb=50 50 554 770,width=9cm, angle=-90]{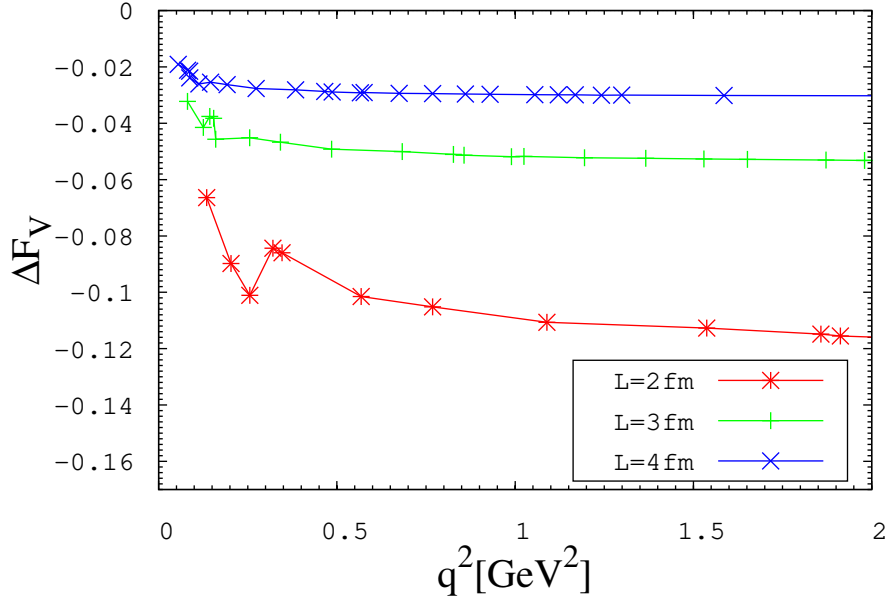}  
     \vspace{-0pt}
     \caption{Numerical estimates for $\Delta F_V$.}
     \label{fig:DeltaF}
  \end{center}
\end{figure}

Our numerical estimates for $\Delta F_V(q_0,{\bf q})$ at $L=T/2=2,3,4\, {\rm fm}$  are presented 
in Fig.~\ref{fig:DeltaF}.
Here, we denote $q^0= i\left( \sqrt{{\bf p}_f^2 + M_\pi^2 } - \sqrt{{\bf p}_i^2 + M_\pi^2 } \right)$,
assuming the dispersion relation of the pion energy, 
${\bf q} = {\bf p}_f - {\bf p}_i$, and choose $M_\pi = 135 {\rm MeV}$,
$F_{\rm eff} = 92.2 {\rm MeV}$, as inputs.
The zig-zag behavior may be due to the lack of the rotational symmetry on the lattice.
Since $F_V^\infty (q^2)$ is an $\mathcal{O}(1)$ quantity, 
our result shows the remaining finite volume effects is around
a few \% already at $L=3\, {\rm fm}$, even when $m_\pi L<1$.


\section{Summary and discussion}
We have studied finite volume effects
on the electro-magnetic pion form factor
in the $\epsilon$ regime.
The pseudoscalar-vector-pseudoscalar three point function
has been calculated in the $\epsilon$ expansion
of chiral perturbation theory to the next-to-leading order.

The dominant finite volume effects, which come from the zero-mode
of the pions can be removed by two simple manipulations:
by inserting non-zero momentum to relevant operators
(or making a subtraction at different time correlators)
and taking a appropriate ratio of them.
After these manipulations, one can safely extract the
electro-magnetic pion form factor for which
the remaining finite volume correction from the non-zero modes
is suppressed to a few percent level already
at $L=3\, {\rm fm}$ even in the $\epsilon$ regime 
(see Figure \ref{fig:DeltaF}). 

It is important to note that our analysis has been done
without using any special features of the $\epsilon$ expansion,
and the dominance of the zero-mode contribution
is expected to be a common feature of finite volume effects in any regime of QCD.
Therefore, our method can be useful for simulations in the $p$ regime,
including the ones with twisted boundary conditions \cite{Mehen:2005fw, Tiburzi:2014yra}.
We also expect a wide application to other quantities like form factors of heavier hadrons.


The authors thanks P. H. Damgaard, S. Hashimoto, T. Onogi, S. Yamaguchi for useful discussions.
The work of HF is supported by the Grant-in-Aid of the Japanese Ministry of Education 
(No. 25800147).

\label{sec:summary}

\appendix
\section{Zero-mode integral}
\label{app:U0integral}

In this appendix, we evaluate the $U_0$ integrals
which are necessary for numerical estimation of
$Z_{M}^{\rm 2pt}$ or $Z_{M}^{\rm 3pt}$.
Although our analysis in this paper is done only in the unquenched QCD,
we use the partially quenched results by 
\cite{Splittorff-Verbaarschot, Fyodorov-Akemann},
because some expressions are simpler for the partially quenched results, 
and the results would be easily extended
to the partially quenched study in these expressions.
The unquenched results are obtained simply by setting 
the valence quark mass $m_v$ to the one of the sea quark masses.

We start with the so-called graded partition function
which consists of $n$ bosons and $m$ fermions.
Its non-perturbative analytic form is given by \cite{Splittorff-Verbaarschot, Fyodorov-Akemann}
\begin{eqnarray}
\mathcal{Z}_{n,m}^Q (\{ \mu_i \})
&=&
\frac{{\rm det} [ \mu_i^{j-1} \mathcal{J}_{Q+j-1} (\mu_i) ]_{i,j=1,\cdots n+m}}
{\prod_{j>i=1}^n (\mu_j^2 - \mu_i^2) \prod_{j>i=n+1}^{n+m} (\mu_j^2 - \mu_i^2)},
\end{eqnarray}
in a fixed topological sector of $Q$.
Here $\mathcal{J}$'s are defined as
$\mathcal{J}_{Q+j-1} (\mu_i) \equiv (-1)^{j-1} K_{Q+j-1}(\mu_i)$
for $i=1,\cdots n$ and
$\mathcal{J}_{Q+j-1} (\mu_i) \equiv  I_{Q+j-1}(\mu_i)$
for $i=n+1,\cdots n+m$,
where $K_\nu$ and $I_\nu$ are the modified Bessel functions.
Partial quenching is completed 
by taking the boson masses to those of valence fermions.

Integrals of some diagonal matrix elements
are obtained by simply differentiating the partition function,
\begin{eqnarray}
\mathcal{S}_v
&\equiv&
\frac{1}{2}
\left\langle \uz_{vv} + \uzd_{vv} \right\rangle_{U_0}
=
\lim_{\mu_b\to\mu_v} \frac{\partial}{\partial\mu_v}
\ln \mathcal{Z}_{1,1+N_f}^Q \left( \mu_b,\mu_v, \{\mu_{sea}\} \right),
\nonumber\\
\mathcal{D}_v
&\equiv&
\frac{1}{4}
\left\langle \left( \uz_{vv} + \uzd_{vv} \right)^2 \right\rangle_{U_0}
\nonumber\\
&=&
\frac{1}{\mathcal{Z}_{N_f}^Q(\{\mu_{sea}\})}
\lim_{\mu_b\to\mu_v}
\frac{\partial^2}{\partial\mu_v^2}
\mathcal{Z}_{1,1+N_f}^Q \left( \mu_b,\mu_v, \{\mu_{sea}\} \right),
\nonumber\\
\mathcal{D}_{v_1v_2}
&\equiv&
\frac{1}{4}
\left\langle \left( \uz_{v_1v_1} +\uzd_{v_1v_1} \right)
\left( \uz_{v_2v_2} +\uzd_{v_2v_2} \right) \right\rangle_{U_0}
\nonumber\\
&=&
\frac{1}{\mathcal{Z}_{N_f}^Q(\{\mu_{sea}\})}
\lim_{\mu_{b_1}\to\mu_{v_1},\mu_{b_2}\to\mu_{v_2}}
\frac{\partial}{\partial\mu_{v_1}}\frac{\partial}{\partial\mu_{v_2}}
\mathcal{Z}_{2,2+N_f}^Q \left( \mu_{b_1},\mu_{b_2},\mu_{v_1},\mu_{v_2}, \{\mu_{sea}\} \right),
\end{eqnarray}
and
\begin{eqnarray}
\mathcal{T}_{v_1v_2}
&\equiv&
\frac{1}{8}
\left\langle \left( \uz_{v_1v_1} + \uzd_{v_1v_1} \right)^2
\left( \uz_{v_2v_2} + \uzd_{v_2v_2} \right)  \right\rangle_{U_0}
\nonumber\\
&=&
\frac{1}{\mathcal{Z}_{N_f}^Q(\{\mu_{sea}\})}
\lim_{\mu_{b_1}\to\mu_{v_1},\mu_{b_2}\to\mu_{v_2}}
\frac{\partial^2}{\partial\mu_{v_1}}\frac{\partial}{\partial\mu_{v_2}}
\mathcal{Z}_{2,2+N_f}^Q \left( \mu_{b_1},\mu_{b_2},\mu_{v_1},\mu_{v_2}, \{\mu_{sea}\} \right).
\end{eqnarray}
Then, $U_0$ integrals
for the degenerate case $m_1=m_2$
can be written as
\begin{eqnarray}
\left\langle \mathcal{B}(U_0) \right\rangle
&=&
2\left[ 1 + \frac{Q^2}{\mu_1} -\frac{2}{N_f}\mathcal{D}_1
+ \left( 1+ \frac{2}{N_f} \right) \mathcal{D}_{11} \right],
\\
\left\langle\mathcal{D}^0(U_0)\right\rangle_{U_0} 
&=& 4\mathcal{S}_{1},\\
\left\langle\mathcal{D}^1(U_0)\right\rangle_{U_0} 
&=& \frac{N_f}{\mu_1}
\left(1-\mathcal{D}_{11}-\frac{Q^2}{\mu_1^2}\right),\\
\left\langle\mathcal{D}^2(U_0)\right\rangle_{U_0} 
&=& -\frac{2}{\mu_1}
\left(\partial_1 \mathcal{S}_{1}-\frac{\mathcal{S}_1}{\mu_1}-\frac{2Q^2}{\mu_1}\mathcal{S}_1
\right),\\
\left\langle\mathcal{D}^3(U_0)\right\rangle_{U_0} 
&=&-\frac{4}{\mu_1}
\left(\frac{1}{N_f}-\frac{\mathcal{D}_{11}}{N_f}-\frac{3Q^2}{N_f\mu_1^2}-\partial_1\mathcal{S}_{1}\right),\\
\left\langle\mathcal{D}^4(U_0)\right\rangle_{U_0} 
&=&\frac{4}{N_f^2}
\partial_1\mathcal{D}_{1}
+
\frac{2}{\mu_1}\partial_1\mathcal{S}_1
+
\frac{4(N_f-2)}{N_f^2}\partial_1\mathcal{D}_{1j}|_{m_j=m_1},
\nonumber\\
\left\langle\mathcal{E}(U_0)\right\rangle_{U_0} 
&=& 2
\left(1+3\mathcal{D}_{11}
+\frac{Q^2}{\mu_1^2}
\right),
\end{eqnarray}
\begin{eqnarray}
\left\langle\mathcal{G}(U_0)\right\rangle_{U_0} 
&=&
2
\left[\mathcal{T}_{11}-\frac{\partial_1\mathcal{D}_{1}}{2}
-\frac{3\mathcal{D}_1}{2\mu_1}
\right.
+\left(
-3+\frac{-4N_f+3}{2\mu_1}
\right)\mathcal{D}_{11}
\nonumber\\&&\hspace{0.3in}
+\left(
3
+\frac{3}{2\mu_1^2}+\frac{3Q^2}{\mu_1^2}\right)\mathcal{S}_{1}
\left.-1-\frac{Q^2}{\mu_1^2}
\left(1+\frac{N_f}{\mu_1}
\right)\right],
\end{eqnarray}
\begin{eqnarray}
\left\langle\mathcal{H}(U_0)\right\rangle_{U_0} 
&=&
-4
\left[\frac{1-\mathcal{D}_{11}}{2N_f\mu_1}
-\frac{\partial_1\mathcal{S}_1}{\mu_1}
-\frac{3Q^2}{2\mu_1^3N_f}\right],
\end{eqnarray}

Here, we have used
\begin{eqnarray}
\lim_{\mu_1\to \mu_2}\frac{\mathcal{S}_1-\mathcal{S}_2}{\mu_1-\mu_2}&=&\partial_1 \mathcal{S}_1
\end{eqnarray}
Note that the derivative $\partial_v$ is taken w.r.t the valence degree of freedom
{\it after} $\mu_b=\mu_v$ limit is taken. 
This partially quenched expression is simpler than that of unquenched theory,
as shown in Ref.~\cite{Damgaard:2007ep}.

It is also useful to note
\begin{eqnarray}
\mathcal{D}_{11}&\equiv& \lim_{\mu_2\to \mu_1}\mathcal{D}_{12}
= - \left.\frac{1}{\mathcal{Z}^Q_{0,N_f}({\mu_{sea}})}
\frac{\partial}{\partial \mu_b}\frac{\partial}{\partial \mu_v}
\mathcal{Z}^Q_{1,1+N_f}(\mu_b,\mu_v,\{\mu_{sea}\})\right|_{\mu_b=\mu_v=\mu_1},
\end{eqnarray}
which was shown in Appendix of Ref~\cite{Aoki:2011pza}.
With this, the following non-trivial relations are obtained,
\begin{eqnarray}
\partial_1 \mathcal{S}_1=\mathcal{D}_1-\mathcal{D}_{11},\;\;\;
\partial_1^2 \mathcal{S}_1=\partial_1\mathcal{D}_1-2\partial_1\mathcal{D}_{12}|_{m_2=m_1}.
\end{eqnarray}
Similarly, we can use 
\begin{eqnarray}
\mathcal{T}_{11}&\equiv& \lim_{\mu_2\to \mu_1}\mathcal{T}_{21}
= - \left.\frac{1}{\mathcal{Z}^Q_{0,N_f}({\mu_{sea}})}
\frac{\partial}{\partial \mu_b}\frac{\partial^2}{\partial^2 \mu_v}
\mathcal{Z}^Q_{1,1+N_f}(\mu_b,\mu_v,\{\mu_{sea}\})\right|_{\mu_b=\mu_v=\mu_1}.
\end{eqnarray}

\section{Loop momentum summations}
\label{app:Iintegral}

In the calculation of the one-loop diagram,
we have encountered the momentum summation:
\begin{eqnarray}
I_{\mu\nu} (q_0,{\bf q}) &=&
\frac{1}{V}
\sum_{p\neq 0, q}\frac{p^\mu(q^\nu-2p^\nu)}{p^2(q-p)^2}
\;\;\; (q^2=q_0^2+{\bf q}^2).
\end{eqnarray}
From the symmetry, on a finite volume $V=TL^3$ we can decompose it as
\begin{eqnarray}
I_{\mu\nu} (q_0,{\bf q}) &=& 
\delta_{\mu \nu}I_1(q_0,{\bf q})
+\delta_{\mu 0} \delta_{\nu 0} I_2(q_0,{\bf q})
+q_\mu q_\nu I_3(q_0,{\bf q}).
\end{eqnarray}
Note that another possible choice $\sum_{i=1}^3\delta_{\mu i} \delta_{\nu i}$
is not independent from the others since
$\delta_{\mu \nu}=\delta_{\mu 0} \delta_{\nu 0}+\sum_{i=1}^3\delta_{\mu i} \delta_{\nu i}$.

For a vector $\bar{q}_\mu$ which satisfy $q\cdot \bar{q}=0,$
we can simplify
\begin{eqnarray}
\label{eq:I_1}
I_{\mu\nu} (q_0,{\bf q}) \bar{q}^\nu &=& \bar{q}_\mu I_1(q_0,{\bf q})+\delta_{\mu0}\bar{q}_0I_2(q_0,{\bf q}).
\end{eqnarray}
In particular, it is useful to note
\begin{eqnarray}
\label{eq:I_1I_2}
I_{0\nu} (q_0,{\bf q}) \bar{q}^\nu &=& \bar{q}_0 l(q_0,{\bf q}),
\end{eqnarray}
where
\begin{eqnarray}
l(q_0,{\bf q})
&\equiv&
I_1(q_0,{\bf q})+I_2(q_0,{\bf q}).
\end{eqnarray}

%
%
%
\end{document}